\documentclass[12pt]{article}

\usepackage{graphicx}
\usepackage{amsfonts}

\def\gtwid{\mathrel{\raise.3ex\hbox{$>$\kern-.75em\lower1ex\hbox{$\sim$}}}}
\def\ltwid{\mathrel{\raise.3ex\hbox{$<$\kern-.75em\lower1ex\hbox{$\sim$}}}}
\def\square{\kern1pt\vbox{\hrule height 1.2pt\hbox{\vrule width 1.2pt\hskip 3pt
   \vbox{\vskip 6pt}\hskip 3pt\vrule width 0.6pt}\hrule height 0.6pt}\kern1pt}

\begin{document}

\begin{titlepage}

\begin{flushright}
UFIFT-QG-15-06 \\
CCTP-2015-15
\end{flushright}

\vskip .5cm

\begin{center}
{\bf Precision Predictions for the Primordial Power Spectra of Scalar Potential Models of Inflation}
\end{center}

\vskip 1cm

\begin{center}
D. J. Brooker$^{1*}$, N. C. Tsamis$^{2\star}$ and R. P. Woodard$^{1\dagger}$
\end{center}

\vskip .5cm

\begin{center}
\it{$^{1}$ Department of Physics, University of Florida,\\
Gainesville, FL 32611, UNITED STATES}
\end{center}

\begin{center}
\it{$^{2}$ Institute of Theoretical Physics \& Computational Physics, \\
Department of Physics, University of Crete, \\
GR-710 03 Heraklion, HELLAS}
\end{center}

\vspace{.5cm}

\begin{center}
ABSTRACT
\end{center}

We exploit a new numerical technique for evaluating the tree order 
contributions to the primordial scalar and tensor power spectra for 
scalar potential models of inflation. Among other things we use the
formalism to develop a good analytic approximation which goes beyond
generalized slow roll expansions in that (1) it is not contaminated 
by the physically irrelevant phase, (2) its 0th order term is exact 
for constant first slow roll parameter, and (3) the correction is 
multiplicative rather than additive. These features allow our 
formalism to capture at first order, effects which are higher order
in other expansions. Although this accuracy is not necessary to 
compare current data with any specific model, our method has a 
number of applications owing to the simpler representation it 
provides for the connection between the power spectra and the
expansion history of a general model.

\begin{flushleft}
PACS numbers: 04.50.Kd, 95.35.+d, 98.62.-g
\end{flushleft}

\vskip .5cm

\begin{flushleft}
$^{*}$ e-mail: djbrooker@ufl.edu \\
$^{\star}$ e-mail: tsamis@physics.uoc.gr \\
$^{\dagger}$ e-mail: woodard@phys.ufl.edu
\end{flushleft}

\end{titlepage}

\section{Introduction}\label{intro}

A central prediction of primordial inflation is the generation of
a nearly scale invariant spectrum of tensor \cite{Starobinsky:1979ty}
and scalar \cite{Mukhanov:1981xt} perturbations. These are increasingly
recognized as quantum gravitational phenomena \cite{Woodard:2009ns,
Ashoorioon:2012kh,Krauss:2013pha}. The scalar power spectra was first
resolved in 1992 \cite{Smoot:1992td} and is by now observed with
stunning accuracy by a variety of ground and space-based detectors \cite{Hinshaw:2012aka,Hou:2012xq,Sievers:2013ica,Ade:2013zuv}. The
inflation community was transfixed with the March 2014 announcement
that BICEP2 had resolved the tensor power spectrum \cite{Ade:2014xna}.
Although subsequent work has shown this signal to be attributable to 
polarized dust emission \cite{Adam:2014bub,Ade:2015tva}, inclusion 
of the BICEP2 data gives the strongest upper bound on the 
tensor-to-scalar ratio \cite{Ade:2015xua}, so we have crossed the
critical threshold at which the tensor power spectrum is more 
accurately constrained by polarization data than by temperature
data. No one knows if the tensor signal is large enough to be resolved
with current technology but many increasingly sensitive polarization 
experiments are planned, under way or actually analysing data, including 
POLARBEAR2 \cite{Hattori:2013jda}, PIPER \cite{Lazear:2014bga}, SPIDER 
\cite{Rahlin:2014rja}, BICEP3 \cite{Ahmed:2014ixy}, and EBEX
\cite{MacDermid:2014wca}.

The triumphal progress of observational cosmology has not seen a comparable 
development of inflation theory. There are many, many theories for what caused 
primordial inflation, and all of them make different predictions for the
tree order power spectra \cite{Mukhanov:1990me,Liddle:1993fq,Lidsey:1995np}.
In most cases there is no precise analytic prediction \cite{Stewart:1993bc}
and numerical techniques must be employed instead \cite{Wang:1997cw,
Martin:1999wa,Easther:2010qz,Mortonson:2010er,Easther:2011yq}. At loop order 
the situation is even worse because there are excellent reasons for doubting
that the naive correlators represent what is being observed 
\cite{Urakawa:2009my,Urakawa:2009gb,Giddings:2010nc,Giddings:2010ui} but
there is no agreement on what should replace them \cite{Byrnes:2010yc,
Urakawa:2010it,Gerstenlauer:2011ti,Giddings:2011zd,Miao:2012xc,Tanaka:2012wi,
Tanaka:2013xe,Miao:2013oko,Tanaka:2014ina}. Although defining loop corrections 
is by no means urgent, it may eventually become relevant with the full 
development of the data on the matter power spectrum which is potentially 
recoverable from highly redshifted 21 cm radiation \cite{Loeb:2003ya,
Furlanetto:2006jb}.

We cannot right now do anything about the multiplicity of models, or about 
the ambiguity in how to define loop corrections for any one of these models. 
Our goal is to instead devise a good analytic prediction for the tree order 
power spectra from any scalar potential model. We will elaborate a numerical
scheme developed previously \cite{Romania:2011ez,Romania:2012tb}, both to
make the scheme even more efficient and to motivate what should be an excellent 
analytic approximation for the power spectra. One fascinating feature of this
formalism is that the tensor power spectrum can easily be converted into its
scalar cousin, so one need only work with the simpler tensor result. We use 
the numerical formalism to examine a wide variety of models with the aim of 
answering two questions:
\begin{enumerate}
\item{For models in which the first slow roll parameter $\epsilon(t)$ evolves, 
what value of the constant $\epsilon$ approximation gives the best fit to the 
actual power spectrum? and}
\item{How numerically accurate is our analytic formula for the nonlocal 
correction factor to the constant $\epsilon$ approximation?}
\end{enumerate}

It is useful to compare and contrast our formalism with the generalized slow 
roll approximation introduced by Stewart \cite{Stewart:2001cd}, and developed 
by Dvorkin and Hu \cite{Dvorkin:2009ne}, to deal with models for which 
$\epsilon(t)$ is small but some of its derivatives are order one when expressed
in Hubble units. In that technique one expands the mode function about its de 
Sitter limit, using the de Sitter Green's function to develop a series of 
nonlocal corrections which depend upon the past history of $\epsilon(t)$. In 
contrast, our formalism is based on the norm-squared of the mode function, 
which avoids having to keep track of the complicated and physically irrelevant 
phase. That allows our first order corrections to recover effects which are
second order in the generalized slow roll approximation. Another difference is 
that our 0th order term is exact for arbitrary constant $\epsilon(t)$. Finally, 
our corrections are multiplicative rather than additive.
  
Although our formalism is more accurate, at the same order, than the 
generalized slow roll expansion, it is debatable whether or not current
observations require greater accuracy for comparison with any specific model. 
Our motivation is rather to better understand how a {\it general} expansion 
history affects the power spectra. This has applications for the power spectra
on three times scales: the interpretation of anomalies in current data; the 
next generation of observations which will reduce the error on $n_s$ by a 
factor of five and might resolve the tensor power spectrum; and in the very 
long term, when the full development of 21 cm cosmology might provide enough 
data to resolve one loop corrections. These applications are:
\begin{itemize}
\item{To facilitate the deconvolution of anomalies in the power spectrum so 
as to identify the sorts of models which might have produced them;}
\item{To generalize the famous single-scalar consistency relation 
\cite{Polarski:1995zn,GarciaBellido:1995fz,Sasaki:1995aw} so one can say 
something even with sparse data, before the tensor spectral index has been
well measured; and}
\item{To understand whether or not loop corrections can receive significant
contributions from early times when $\epsilon(t)$ was small and $H(t)$
was large.}
\end{itemize}
Regarding the 3rd point, one should note that the $\zeta$-$\zeta$ propagator
contains a factor of $1/\epsilon(t)$ which is usually assumed to be cancelled
by powers of $\epsilon(t)$ from the vertices \cite{Woodard:2014jba}. However,
it seems possible that the propagator --- which depends nonlocally on 
$\epsilon(t)$ --- might receive significant contributions from small, early
values of $\epsilon(t)$. If so, one might expect loop corrections from vertices
at late times to be enhanced by large factors of $\epsilon_{\rm late}/
\epsilon_{\rm early}$. A closely related issue is deciding what time best 
describes the putative loop counting parameter of $G H^2(t)$ 
\cite{Woodard:2014jba}. 

A different sort of application concerns nonlocal modified gravity models 
of cosmology which are conjectured to represent quantum gravitational effects 
that became nonperturbatively strong during primordial inflation 
\cite{Tsamis:2009ja,Tsamis:2010ph,Romania:2012av}. These quantum gravitational
effects derive from secular growth in the graviton propagator which is 
known for de Sitter \cite{Vilenkin:1982wt,Linde:1982uu,Starobinsky:1982ee},
but not for geometries in which $\epsilon(t)$ evolves \cite{Dolgov:2005se}. 
Our formalism will facilitate better extrapolations of these growth factors to 
general geometries, which should motivate more realistic models.

This paper consists of six sections, of which the first is this Introduction.
Section 2 reviews scalar potential models and the simple procedure for passing
from the expansion history to the potential and vice versa. In section 3 we 
define the two tree order power spectra, explain the relation between them, 
and give constant $\epsilon$ results. Our improved formalism is derived in
section 4, along with the analytic approximation. Section 5 presents numerical
studies. Our conclusions comprise section 6.

\section{Scalar Potential Models}\label{single}

The Lagrangian for a general scalar potential model is,
\begin{equation}
\mathcal{L} = \frac1{16 \pi G} \, R \sqrt{-g} - \frac12 
\partial_{\mu} \varphi \partial_{\nu} \varphi g^{\mu\nu}
\sqrt{-g} - V(\varphi) \sqrt{-g} \; .
\end{equation}
We assume a homogeneous, isotropic and spatially flat 
background characterized by $\varphi_0(t)$ and scale
factor $a(t)$,
\begin{equation}
ds^2 = -dt^2 + a^2(t) d\vec{x} \!\cdot\! d\vec{x} \qquad 
\Longrightarrow \qquad H(t) \equiv \frac{\dot{a}}{a} \quad , 
\quad \epsilon(t) \equiv -\frac{\dot{H}}{H^2} \; .
\end{equation}
The nontrivial Einstein equations for this background are,
\begin{eqnarray}
3 H^2 & = & 8\pi G \Bigl[ \frac12 \dot{\varphi}_0^2 + 
V(\varphi_0)\Bigr] \; , \label{E1} \\
-2\dot{H} - 3 H^2 & = & 8\pi G \Bigl[ \frac12 
\dot{\varphi}_0^2 - V(\varphi_0) \Bigr] \; . \label{E2}
\end{eqnarray}

As long as the tensor power spectrum remains unresolved
there is no question that scalar potential models can be 
devised to fit the data because one can regard the 
observed scalar power spectrum as a first order
differential equation for $H(t)$ \cite{Salopek:1988qh}.
Once $H(t)$ is known there is a simple way of using 
equations (\ref{E1}-\ref{E2}) to construct a potential 
$V(\varphi)$ which supports any function $H(t)$ that
obeys $\dot{H}(t) < 0$ \cite{Tsamis:1997rk,Saini:1999ba,
Capozziello:2005mj,Woodard:2006nt,Guo:2006ab}. One first 
adds (\ref{E1}) and (\ref{E2}) to obtain an equation for 
the scalar background,
\begin{equation}
\varphi_0(t) = \varphi_i \pm \int_{t_i}^{t} \!\! dt'
\sqrt{\frac{-\dot{H}(t')}{16 \pi G}} \; . \label{E3}
\end{equation}
By graphing this relation and then rotating the graph by
$90^{\circ}$ one can easily invert (\ref{E3}) to solve 
for time $t(\varphi)$. The final step is to subtract
(\ref{E2}) from (\ref{E1}) to find the potential,
\begin{equation}
V(\varphi) = \frac1{8\pi G} \Bigl[ \dot{H}(t) \!+\! 3
H^2(t)\Bigr]_{t = t(\varphi)} \; . \label{E4}
\end{equation}

Rather than specifying the expansion history $a(t)$ and
using relations (\ref{E3}-\ref{E4}) to reconstruct the
potential, it is more usual to specify the potential and
then solve for the expansion history $a(t)$. This is 
greatly facilitated by making the slow roll approximation,
\begin{equation}
H \approx \sqrt{\frac13 8\pi G V(\varphi)} \qquad , \qquad
\epsilon \approx \frac1{16\pi G} \Bigl[ \frac{V'(\varphi)}{
V(\varphi)}\Bigr]^2 \; . \label{slowroll}
\end{equation}
It is desirable to express the scale factor $a = a_i e^N$
in terms of the number of e-foldings $N$ from the beginning 
of inflation. Then one can use the slow roll approximation
(\ref{slowroll}) to solve for the scalar's evolution from
initial value $\varphi_i$ by inverting the relation,
\begin{equation}
N = -8\pi G \!\! \int_{\varphi_i}^{\varphi} \!\!\!\! d\psi
\frac{V(\psi)}{V'(\psi)} \; . \label{Nfromphi}
\end{equation}
An important special case is power law potentials,
\begin{equation}
V(\varphi) = A \varphi^{\alpha} \qquad \Longrightarrow \qquad
\epsilon = \frac{\epsilon_i}{1 \!-\! \frac{4 \epsilon_i}{\alpha}
\, N} \quad , \quad H = H_i \Bigl[ 1 \!-\! \frac{4 \epsilon_i}{
\alpha} \, N\Bigr]^{\frac{\alpha}{4}} \; . \label{powerlaw}
\end{equation}

\section{The Primordial Power Spectra}\label{power}

The purpose of this section is to introduce notation to describe the
scalar and tensor power spectra and review the local approximate formulae
for them. We begin by defining the two spectra, and explaining how the 
tensor result can be used to derive the scalar result. Then we consider
the special cases of expansion histories with constant $\epsilon(t)$, 
and where $\epsilon(t)$ makes an instantaneous transition from one 
constant value of $\epsilon(t)$ to another.

\subsection{Generalities}

It is useful to define time dependent extensions of the scalar and tensor
power spectra, $\Delta^2_{\mathcal{R}}(k)$ and $\Delta^2_h(k)$. At tree
order these time dependent power spectra take the form of constants times
the norm-squared of the scalar and tensor mode functions $v(t,k)$ and 
$u(t,k)$,
\begin{eqnarray}
\Delta^2_{\mathcal{R}}(t,k) & = & \frac{k^3}{2 \pi^2} \!\times\! 4\pi G 
\!\times\! \vert v(t,k)\vert^2 \; , \label{scalarpower} \\
\Delta^2_{h}(t,k) & = & \frac{k^3}{2 \pi^2} \!\times\! 32 \pi G \!\times\!
2 \!\times\! \vert u(t,k) \vert^2 \; . \label{tensorpower}
\end{eqnarray}
The actual primordial power spectra are defined by evaluating these time
dependent ones long after the time $t_k$ of first horizon crossing at
which $k = H(t_k) a(t_k)$. After $t_k$ the mode functions approach
constants, and it is these constant values which define the predicted 
power spectra,
\begin{equation}
\Delta^2_{\mathcal{R}}(k) \equiv \Delta^2_{\mathcal{R}}(t,k) 
\Bigl\vert_{t \gg t_k} \qquad , \qquad \Delta^2_{\mathcal{R}}(k) \equiv 
\Delta^2_{\mathcal{R}}(t,k) \Bigl\vert_{t \gg t_k} \; . \label{latetime}
\end{equation}

The equations of motion and normalization conditions for the scalar and 
tensor mode functions are,
\begin{eqnarray}
\ddot{v} + \Bigl(3 H \!+\! \frac{\dot{\epsilon}}{\epsilon} \Bigr) \dot{v}
+ \frac{k^2}{a^2} v = 0 \quad & , & \quad v \dot{v}^* - \dot{v} v^* = 
\frac{i}{\epsilon a^3} \; , \label{scalareqn} \\
\ddot{u} + 3 H \dot{u} + \frac{k^2}{a^2} u = 0 \quad & , & \quad 
u \dot{u}^* - \dot{u} u^* = \frac{i}{a^3} \; . \label{tensoreqn} 
\end{eqnarray}
The full system (\ref{scalarpower}-\ref{tensoreqn}) is frustrating because
the phenomenological predictions (\ref{latetime}) emerge from late times
whereas it is only at early times $k \gg H(t) a(t)$ at which one has a
good asymptotic form for the mode functions,
\begin{eqnarray}
v(t,k) & {\mbox{} \atop \overrightarrow{\mbox{\tiny $k \gg H a$} }} & 
\frac1{\sqrt{2 k \epsilon(t) a^2(t)}} \, \exp\Bigl[ -i k \!\! 
\int_{t_i}^{t} \!\!\! \frac{dt'}{a(t')} \Bigr] \; , \label{WKBv} \\
u(t,k) & {\mbox{} \atop \overrightarrow{\mbox{\tiny $k \gg H a$}}} & 
\frac1{\sqrt{2 k a^2(t)}} \, \exp\Bigl[ -i k \!\! \int_{t_i}^{t} \!\!\!
\frac{dt'}{a(t')} \Bigr] \; . \label{WKBu}
\end{eqnarray}
So these forms (\ref{WKBv}-\ref{WKBu}) serve to define the initial 
conditions, and one must then use equations (\ref{scalareqn}-\ref{tensoreqn})
to evolve $v(t,k)$ and $u(t,k)$ forward until well past first horizon 
crossing, at which point the mode functions are nearly constant and one
can use them in expressions (\ref{scalarpower}-\ref{latetime}) to compute
the primordial power spectra. It is this cumbersome and highly model dependent
procedure which we seek to simplify and systematize.

First, we take note of an important relation between the scalar and
tensor systems. This is that the scalar relations (\ref{scalareqn}) follow 
from the tensor ones (\ref{tensoreqn}) by simple changes of the scale factor 
and time \cite{Tsamis:2003zs},
\begin{equation}
a(t) \longrightarrow \sqrt{\epsilon(t)} \!\times\! a(t) \qquad , \qquad
\frac{\partial}{\partial t} \longrightarrow \frac1{\sqrt{\epsilon(t)}}
\!\times\! \frac{\partial}{\partial t} \; . \label{MtoN}
\end{equation}
We will therefore concentrate on the tensor system, and we do so in
terms of the norm-squared tensor mode function,
\begin{equation}
M(t,k) \equiv \vert u(t,k) \vert^2 \; .
\end{equation}

\subsection{The case of constant $\epsilon(t)$}

An important special case is when $\epsilon(t)$ is constant, for which
the Hubble parameter and scale factor are,
\begin{equation}
\epsilon(t) = \epsilon_i \qquad \Longrightarrow \qquad H(t) = 
\frac{H_i}{1 \!+\! \epsilon_i H_i \Delta t} \quad , \quad a(t) =
\Bigl[1 \!+\! \epsilon_i H_i \Delta t \Bigr]^{\frac1{\epsilon_i}} \; ,
\end{equation}
where $\Delta t \equiv t - t_i$. Note that the combination $H(t) [
a(t)]^{\epsilon}$ is constant. The appropriate tensor mode
function for constant $\epsilon(t)$ is,
\begin{equation}
u_0(t,k) = \frac1{a(t) \sqrt{2k}} \times \sqrt{ \frac{\pi z}{2}} 
H^{(1)}_{\nu}(z) \;\; , \;\; z(t,k) \equiv \frac{k}{(1 \!-\! \epsilon) 
H a} \;\; , \;\; \nu \equiv \frac32 + \frac{\epsilon}{1 \!-\! \epsilon} 
\; . \label{consteu}
\end{equation}
From the small argument expansion of the Hankel function we can infer
the constant late time limit of (\ref{consteu}),
\begin{eqnarray}
u_0(t,k) & {\mbox{} \atop \overrightarrow{\mbox{\tiny $k \ll H a$}}} &
\sqrt{ \frac{\pi z}{4 a^2 k}} \!\times\! -\frac{i \Gamma(\nu)}{\pi}
\Bigl( \frac{2}{z}\Bigr)^{\nu} \; , \\
& = & -\frac{i (1 \!+\! \epsilon) \Gamma(\frac12 \!+\! \frac{\epsilon}{1-
\epsilon})}{\sqrt{2 \pi k^3}} \Bigl[2(1 \!-\! \epsilon)\Bigr]^{
\frac{\epsilon}{1-\epsilon}} \Bigl[\frac{H(t) a^{\epsilon}(t)}{k^{\epsilon}}\Bigr]^{\frac1{1-\epsilon}} \; . \label{u0late}
\end{eqnarray}
It is usual to evaluate the constant factor of $H(t) a^{\epsilon}(t)$ at
horizon crossing,
\begin{equation}
u_0(t,k) \quad \overrightarrow{\mbox{\tiny $k \ll H a$}} \quad
\frac{H(t_k)}{\sqrt{2 k^3}} \!\times\! -\frac{i (1 \!+\! \epsilon)
\Gamma(\frac12 \!+\! \frac{\epsilon}{1 - \epsilon})}{\sqrt{\pi}}
\Bigl[ 2 (1 \!-\! \epsilon)\Bigr]^{\frac{\epsilon}{1-\epsilon}} \; .
\label{bestlateu0}
\end{equation}
Substituting (\ref{bestlateu0}) in expressions 
(\ref{scalarpower}-\ref{latetime}) gives the famous constant $\epsilon$
predictions for the power spectra,
\begin{eqnarray}
\Delta^2_{\mathcal{R}}(k) \Bigl\vert_{\dot{\epsilon}=0} & = & 
\Bigl(\frac{\hbar}{c^5}\Bigr) \!\times\! \frac{G H^2(t_k)}{\pi
\epsilon} \!\times\! \frac{(1 \!+\! \epsilon)^2 \Gamma^2(\frac12 \!+\!
\frac{\epsilon}{1 - \epsilon})}{\pi} \Bigl[ 2 (1 \!-\! \epsilon)
\Bigr]^{\frac{2\epsilon}{1-\epsilon}} \; , \label{scalarconsteps} \\
\Delta^2_{h}(k) \Bigl\vert_{\dot{\epsilon}=0} & = & 
\Bigl(\frac{\hbar}{c^5}\Bigr) \!\times\! \frac{16}{\pi} G H^2(t_k) 
\!\times\! \frac{(1 \!+\! \epsilon)^2 \Gamma^2(\frac12 \!+\!
\frac{\epsilon}{1 - \epsilon})}{\pi} \Bigl[ 2 (1 \!-\! \epsilon)
\Bigr]^{\frac{2\epsilon}{1-\epsilon}} \; . \label{tensorconsteps}
\end{eqnarray}

\begin{figure}[ht]
\includegraphics[width=10.0cm,height=8.0cm]{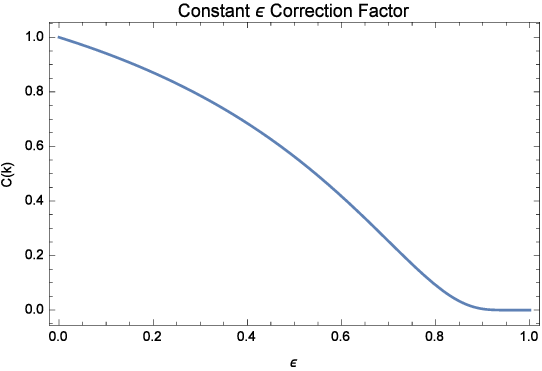}
\caption{Graph of $\frac1{\pi} (1 \!+\! \epsilon)^2 \Gamma^2(\frac12 
\!+\! \frac{\epsilon}{1-\epsilon}) [2 (1 \!-\! \epsilon)]^{
\frac{2\epsilon}{1-\epsilon}}$ as a function of $\epsilon$.}
\label{epsilonfactor}
\end{figure}

The final factor in expressions 
(\ref{scalarconsteps}-\ref{tensorconsteps}) contains an 
$\epsilon$-dependent correction which is not usually quoted
because it is so near unity for small $\epsilon$,
\begin{equation}
C(\epsilon) \equiv \frac{(1 \!+\! \epsilon)^2 \Gamma^2(\frac12 \!+\!
\frac{\epsilon}{1 - \epsilon})}{\pi} \Bigl[ 2 (1 \!-\! \epsilon)
\Bigr]^{\frac{2\epsilon}{1-\epsilon}} \; . \label{epscor}
\end{equation}
Figure~\ref{epsilonfactor} shows the dependence of $C(\epsilon)$
versus $\epsilon$ for the full inflationary range of $0 \leq \epsilon
< 1$. Note that $C(\epsilon)$ is a monotonically decreasing function
of $\epsilon$. In particular, it goes to zero for $\epsilon 
\rightarrow 1^-$. If we assume the single-scalar relation of $r = 16 
\epsilon$ then the current upper bound of $r < 0.09$ implies $\epsilon 
< 0.0056$. At this upper bound the constant $\epsilon$ correction 
factor is about 0.997. It would be even closer to unity for smaller 
vales of $\epsilon$.

\subsection{The case of a jump from $\epsilon(t) = \epsilon_1$ to
$\epsilon(t) = \epsilon_2$}

Suppose the Universe begins with constant $\epsilon(t) = \epsilon_1$,
with initial values of the Hubble parameter and scale factor $H_1$ and
$a_1$, respectively. At some time $t_2$ the first slow roll parameter
makes an instantaneous transition to $\epsilon(t) = \epsilon_2 > 
\epsilon_1$. In both regions we express the scale factor in terms of
the number of e-foldings $N$ as $a(t) = a_1 e^N$. If the transition 
time $t = t_2$ corresponds to $N = N_2$ then we have,
\begin{eqnarray}
N < N_2 & \Longrightarrow & \epsilon = \epsilon_1 \quad , \quad
H = H_1 e^{-\epsilon_1 N} \; , \\
N > N_2 & \Longrightarrow & \epsilon = \epsilon_2 \quad , \quad
H = H_1 e^{\Delta \epsilon N_2 -\epsilon_2 N} \; ,
\end{eqnarray}
where $\Delta \epsilon \equiv \epsilon_2 - \epsilon_1$. 

It is useful to define mode functions assuming the two constant values 
of $\epsilon(t) = \epsilon_i$ had held for all time,
\begin{equation}
u_i(t,k) \equiv \frac1{\sqrt{2 k a^2(t)}} \sqrt{ \frac{\pi z}{2}} \, 
H^{(1)}_{\nu_i}(z) \quad , \quad z \equiv \frac{k}{(1 \!-\! \epsilon_i)
H a} \quad , \quad \nu_i \equiv \frac12 \Bigl( \frac{3 \!-\! \epsilon_i}{1
\!-\! \epsilon_i}\Bigr) \; .
\end{equation}
The actual mode function after the transition is a linear combination of
the positive and negative frequency solutions,
\begin{eqnarray}
N < N_2 & \Longrightarrow & u(t,k) = u_1(t,k) \; , \\
N > N_2 & \Longrightarrow & u(t,k) = \alpha u_2(t,k) + \beta u_2^*(t,k)
\; . \label{uafter}
\end{eqnarray}
The combination coefficients are,
\begin{eqnarray}
\alpha & \!\!\! = \!\!\! & 
\frac{i\pi}{4} \Biggl[ \sqrt{z_1} H^{(1)}_{\nu_1}(z_1)
\Bigl[ \sqrt{z_2} H^{(1)}_{\nu_2}(z_2)\Bigr]^*_{,z_2} \!\!- 
\Bigl[ \sqrt{z_1} H^{(1)}_{\nu_1}(z_1) \Bigr]_{,z_1} 
\sqrt{z_2} H^{(1)*}_{\nu_2}(z_2)\Biggr] , \qquad \label{alpha} \\
\beta & \!\!\! = \!\!\! & 
\frac{i\pi}{4} \Biggl[ -\sqrt{z_1} H^{(1)}_{\nu_1}(z_1)
\Bigl[ \sqrt{z_2} H^{(1)}_{\nu_2}(z_2)\Bigr]_{,z_2} \!\!+ 
\Bigl[ \sqrt{z_1} H^{(1)}_{\nu_1}(z_1) \Bigr]_{,z_1} 
\sqrt{z_2} H^{(1) }_{\nu_2}(z_2)\Biggr] , \qquad \label{beta}
\end{eqnarray}
where $z_1$ and $z_2$ are,
\begin{equation}
z_i \equiv \frac1{1 \!-\! \epsilon_i} \frac{k}{H(t_2) a(t_2)} = 
\frac1{1 \!-\! \epsilon_i} \frac{H(t_k) a(t_k)}{H(t_2) a(t_2)} \; .
\end{equation} 

We seek to understand the effect of varying the transition point $N_2$ 
relative to first horizon crossing $N_k$, with the important dimensional
parameters $k$ and $H(t_k)$ held fixed. Of course this is accomplished
by adjusting the initial values $a_1$ and $H_1$,
\begin{eqnarray}
N_k < N_2 & \Longrightarrow & H_1 = e^{\epsilon_1 N_k} H(t_k) \quad , 
\quad a_1 = \frac{k}{e^{N_k} H(t_k)} \; , \\
N_k > N_2 & \Longrightarrow & H_1 = e^{\epsilon_2 N_k - \Delta \epsilon N_2}
H(t_k) \quad , \quad a_1 = \frac{k}{e^{N_k} H(t_k)} \; .
\end{eqnarray}
It is useful to express the late time limit of $M(t,k)$ in terms of the
results $M_i$ which would pertain if $\epsilon(t) = \epsilon_i$ for all
time,
\begin{equation}
M_i \equiv \frac{H^2(t_k)}{2 k^3} \times C(\epsilon_i) \; ,
\end{equation}
where expression (\ref{epscor}) gives the constant $\epsilon$ correction 
factor $C(\epsilon)$. The late time limit of the actual mode function
$u(t,k)$ always derives from (\ref{uafter}), but the late time limit of
$u_2(t,k)$ depends upon whether the transition comes before or after first
horizon crossing,
\begin{eqnarray}
N_k < N_2 & \Longrightarrow & \lim_{t \rightarrow \infty} u_2(t,k) =
-i \sqrt{M_2} \!\times\! e^{-\frac{\Delta \epsilon}{1 - \epsilon_2} 
(N_k - N_2)} \; , \\
N_k > N_2 & \Longrightarrow & \lim_{t \rightarrow \infty} u_2(t,k) =
-i \sqrt{M_2} \; .
\end{eqnarray}
Hence the late time limit of $M(t,k) = \vert u(t,k)\vert^2$ is,
\begin{eqnarray}
N_k < N_2 & \Longrightarrow & \lim_{t \rightarrow \infty} M(t,k) = 
\vert \alpha \!-\! \beta\vert^2 \!\times\! e^{-\frac{2 \Delta \epsilon}{
1-\epsilon_2} (N_k - N_2)} \!\times\! M_2 \; , \label{DNneg} \\
N_k > N_2 & \Longrightarrow & \lim_{t \rightarrow \infty} M(t,k) =
\vert \alpha \!-\! \beta \vert^2 \!\times\! M_2 \; . \label{DNpos}
\end{eqnarray}
Because only the difference of (\ref{alpha}-\ref{beta}) enters the late 
time limit, the imaginary part of $H^{(1)}_{\nu_2}(z_2) = J_{\nu_2}(z_2)
+ i N_{\nu_2}(z_2)$ drops out,
\begin{equation}
\alpha \!-\! \beta = \frac{i\pi}{2} \Biggl[ \sqrt{z_1} \, 
H^{(1)}_{\nu_1}(z_1) \Bigl[ \sqrt{z_2} \, J_{\nu_2}(z_2)\Bigr]_{, z_2} 
\!- \Bigl[\sqrt{z_1} \, H^{(1)}_{\nu_1}(z_1) \Bigr]_{ , z_1} \sqrt{z_2} 
\, J_{\nu_2}(z_2) \Biggr] . \label{alpabeta}
\end{equation}
In evaluating the $z_i$ one must distinguish between the cases for which 
first horizon crossing occurs before and after the transition,
\begin{eqnarray}
N_k < N_2 & \Longrightarrow & z_i = \frac1{1 \!-\! \epsilon_i}
\, e^{(1 -\epsilon_1) (N_k - N_2)} \; , \\
N_k > N_2 & \Longrightarrow & z_i = \frac1{1 \!-\! \epsilon_i}
\, e^{(1 -\epsilon_2) (N_k - N_2)} \; .
\end{eqnarray}

\begin{figure}[ht]
\includegraphics[width=10.0cm,height=8.0cm]{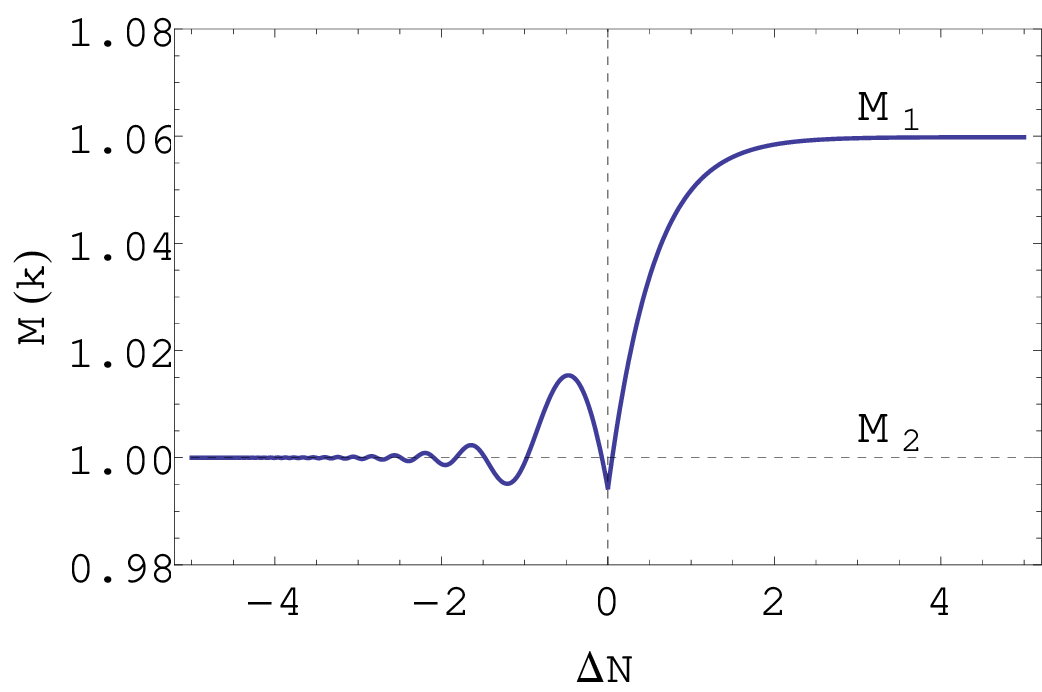}
\caption{Graph of $M(k) = \lim_{t \rightarrow \infty} M(t,k)$ in units 
of $M_2$ for an instantaneous transition from $\epsilon_1 = \frac1{200}$ 
to $\epsilon_2 = \frac1{10}$ as a function of the number of e-foldings 
$\Delta N \equiv N_2 - N_k$ from first horizon crossing. The curve has a 
cusp at $\Delta N = 0$ because $H(t_k)$ and $k$ are held fixed, whereas 
the way they depend upon the initial values of $H(t)$ and $a(t)$ changes 
from $\Delta N < 0$ to $\Delta N > 0$.}
\label{jump}
\end{figure}

Figure~\ref{jump} shows the late time limit of $M(t,k)$ for an 
instantaneous transition from $\epsilon_1 = \frac1{200}$ to 
$\epsilon_2 = \frac1{10}$ at $N = N_2$. For $\Delta N \equiv N_2 - N_k 
\ll -1$ the transition occurs long before first horizon crossing 
so $M(t,k)$ approaches $M_2$, the result for a universe which has
had $\epsilon(t) = \epsilon_2$ for all time. This follows from our
analytic expressions because $z_i \gg 1$ in this regime, so we have,
\begin{eqnarray}
N_k \ll N_2 & \Longrightarrow & \sqrt{ \frac{\pi z_1}{2}} 
H^{(1)}_{\nu_1}(z_1) \longrightarrow \exp\Bigl[ i z_1 \!-\! i
\Bigl( \nu_1 \!+\! \frac12\Bigr) \frac{\pi}2\Bigr] \; , \\
& \Longrightarrow & \sqrt{ \frac{\pi z_2}{2}} J_{\nu_2}(z_2) 
\longrightarrow \cos\Bigl[ z_2 \!-\! \Bigl( \nu_2 \!+\! \frac12\Bigr) 
\frac{\pi}2\Bigr] \; , \\
& \Longrightarrow & \alpha \!-\! \beta \longrightarrow 
\exp\Bigl[i (z_1 \!-\! z_2) \!+\! (\nu_2 \!-\! \nu_1) \frac{\pi}2\Bigr]
\; .
\end{eqnarray}
For $\Delta N \gg +1$ the transition occurs long after first horizon 
crossing, which implies that the new value of $\epsilon(t) = \epsilon_2$ 
is irrelevant and $M(t,k)$ freezes in at the value $M_1$ that would 
pertain for a universe with $\epsilon(t) = \epsilon_1$ for all time. 
This is the regime of $z_i \ll 1$, for which our analytic expressions
give,
\begin{eqnarray}
N_k \ll N_2 & \Longrightarrow & \sqrt{ \frac{\pi z_1}{2}} 
H^{(1)}_{\nu_1}(z_1) \longrightarrow -\frac{\Gamma(\nu_1)}{\sqrt{\pi}}
\Bigl( \frac2{z_1}\Bigr)^{\nu_1 - \frac12} \; , \\
& \Longrightarrow & \sqrt{ \frac{\pi z_2}{2}} J_{\nu_2}(z_2) 
\longrightarrow \frac{\sqrt{\pi}}{\Gamma(1 \!+\! \nu_2)} 
\Bigl( \frac{z_2}{2} \Bigr)^{\nu_2 + \frac12} \; , \\
& \Longrightarrow & \alpha \!-\! \beta \longrightarrow 
\sqrt{\frac{M_1}{M_2}} \exp\Bigl[\frac{\Delta \epsilon (N_k \!-\!
N_2)}{1 \!-\! \epsilon_2} \Bigr] \; .
\end{eqnarray}

Although the details depend upon the values of $\epsilon_1$ and 
$\epsilon_2$, Figure~\ref{jump} really is generic and has been known
since the 1992 study of Starobinsky \cite{Starobinsky:1992ts}. In 
particular, as $N_2$ approaches $N_k$ from below there are always 
oscillations of decreasing frequency and increasing amplitude, the 
value at $N_2 = N_k$ always is somewhat below $M_2 < M_1$, and the 
value for $N_2 > N_k$ always rises monotonically to approach $M_1$. 
Similar results pertain for transitions of the inflaton potential 
\cite{Mukhanov:1991rp,Adams:2001vc}.

\section{Our Evolution Equation}\label{evolve}

This is the main analytic portion of the paper. It begins by reviewing
the derivation of an evolution equation for $M(t,k) \equiv \vert
u(t,k) \vert^2$ \cite{Romania:2011ez,Romania:2012tb}. We then factor 
out the main effect by writing $M(t,k) = M_0(t,k) \times \Delta M(t,k)$, 
where $M_0(t,k) \equiv \vert u_0(t,k) \vert^2$ is the constant $\epsilon$
result evaluated at the instantaneous $\epsilon(t)$. Next $M_0(t,k)$ is 
simplified and the asymptotic behaviors are discussed. By linearizing 
the equation for $\Delta M(t,k)$ we derive what should be an excellent 
approximation for $\Delta M(t,k)$ for a general inflationary expansion 
history. 

\subsection{An evolution equation for $M(t,k)$}

The tensor power spectrum (\ref{tensorpower}) depends upon the 
norm-squared of the tensor mode function $u(t,k)$. It is numerically 
wasteful to follow the irrelevant phase using the tensor evolution 
equations (\ref{tensoreqn}), especially during the early time regime of
$k \gg H(t) a(t)$ when oscillations are rapid. The better strategy is 
to use (\ref{tensoreqn}) to derive an equation for $M(t,k) \equiv \vert
u(t,k) \vert^2$ directly. This is accomplished by computing the first
two time derivatives, 
\begin{eqnarray}
\dot{M}(t,k) & = & u(t,k) \!\times\! \dot{u}^*(t,k) + \dot{u}(t,k) 
\!\times\! u^*(t,k) \; , \label{Mdot} \\
\ddot{M}(t,k) & = & u(t,k) \!\times\! \ddot{u}^*(t,k) + 2 \dot{u}(t,k)
\!\times\! \dot{u}^*(t,k) + \ddot{u}(t,k) \!\times\! u^*(t,k) \; .
\qquad \label{Mddot}
\end{eqnarray}
Now use (\ref{tensoreqn}) to eliminate $\ddot{u}$ and $\ddot{u}^*$ in
(\ref{Mddot}),
\begin{equation}
\ddot{M} = -3 H \dot{M} - \frac{2 k^2}{a^2} M + 2 \dot{u} \dot{u}^* \; .
\label{Mddot1}
\end{equation}
Squaring (\ref{Mdot}) and subtracting the square of the Wronskian 
(\ref{tensoreqn}) gives $\dot{u} \dot{u}^*$,
\begin{eqnarray}
\dot{M}^2 & = & +u^2 \dot{u}^{*2} + 2 M \dot{u} \dot{u}^* + \dot{u}^2
u^{*2} \; , \label{Mdotsq} \\
\frac1{a^6} & = & -u^2 \dot{u}^{*2} + 2 M \dot{u} \dot{u}^* - \dot{u}^2
u^{*2} \; . \label{Wronsq}
\end{eqnarray}
Hence the desired evolution equation for $M(t,k)$ is \cite{Romania:2011ez,
Romania:2012tb},
\begin{equation}
\ddot{M} + 3 H \dot{M} + \frac{2 k^2}{a^2} M = \frac1{2 M} \Bigl[ 
\dot{M}^2 + \frac1{a^6}\Bigr] \; . \label{Meqn}
\end{equation}
As already noted, the transformation (\ref{MtoN}) converts (\ref{Meqn})
into an equation for the norm-squared of the scalar mode function
$N(t,k) \equiv \vert v(t,k) \vert^2$, so both power spectra follow from
$M(t,k)$.

One indication of how much more efficient it is to evolve (\ref{Meqn})
than (\ref{tensoreqn}) comes from comparing the asymptotic expansions
of $u(t,k)$ and $M(t,k)$ in the early time regime of $k \gg H(t) a(t)$.
The expansion for $u(t,k)$ is in powers of $1/k$ and is not even local
at first order,
\begin{eqnarray}
u(t,k) & = & \Biggl\{1 + \frac{i \alpha(t)}{k} + \frac{\beta(t)}{k^2}
+ O\Bigl( \frac1{k^3}\Bigr) \Biggr\} \!\times\! \frac{ \exp[-i k \!
\int_{t_i}^{t} \! \frac{dt'}{a(t')}]}{\sqrt{2 k a^2(t)}} \; , 
\label{uexp} \\
\alpha(t) & = & \frac12 \! \int_{t_i}^{t} \!\! dt' \Bigl[2 \!-\! 
\epsilon(t')\Bigr] H^2(t') a(t') \; , \label{ualpha} \\
\beta(t) & = & -\frac12 \alpha^2(t) + \frac14 \Bigl[2 \!-\! \epsilon(t)
\Bigr] H^2(t) a^2(t) \; . \label{ubeta}
\end{eqnarray}
In contrast, $M(t,k)$ gives a series in $1/k^2$ which is local to all 
orders,
\begin{eqnarray}
M(t,k) & = & \Biggl\{1 + \frac{\overline{\alpha}(t)}{k^2} + 
\frac{\overline{\beta}(t)}{k^4} + O\Bigl(\frac1{k^6}\Bigr) \Biggr\} 
\!\times\! \frac1{2 k a^2(t)} \; , \label{Mexp} \\
\overline{\alpha}(t) & = & \Bigl(1 \!-\! \frac12 \epsilon
\Bigr) H^2 a^2 \; , \label{alphaM} \\
\overline{\beta}(t) & = & \Biggl[ \frac94 \epsilon \Bigl(1 \!-\! \frac23 
\epsilon\Bigr) \Bigl( 1 \!-\! \frac12 \epsilon\Bigr) + \frac{9 
\dot{\epsilon}}{8 H} - \frac{3  \epsilon \dot{\epsilon}}{4 H}
+ \frac{\ddot{\epsilon}}{8 H^2} \Biggr] H^4 a^4 \; . \label{betaM}
\end{eqnarray}
Taking the norm-squared of (\ref{uexp}) helps to explain why our formalism 
is so much more accurate than the generalized slow roll approximation 
\cite{Stewart:2001cd,Dvorkin:2009ne},
\begin{eqnarray}
\Bigl\vert u(t,k)\Bigr\vert^2 & = & \Biggl\{ 1 \!+\! 
\frac{i\alpha}{k} \!+\! \frac{\beta}{k^2} \!+\! O\Bigl(\frac1{k^3}\Bigr)
\Biggr\} \Biggl\{1 \!-\! \frac{i\alpha}{k} \!+\! \frac{\beta}{k^2} \!+\!
O\Bigl(\frac1{k^3}\Bigr) \Biggr\} \!\times\! \frac1{2 k a^2} \; , 
\qquad \\
& = & \Biggl\{ 1 \!+\! \frac{[ \alpha^2 \!+\! 2\beta]}{k^2} \!+\!
O\Bigl(\frac1{k^4}\Bigr) \Biggr\} \!\times\! \frac1{2 k a^2} \; .
\label{badcomp}
\end{eqnarray}
Comparing (\ref{badcomp}) with (\ref{Mexp}) reveals that one must expand 
$u(t,k)$ to second order to recover the first order correction to $M(t,k)$.
Dvorkin and Hu have noted (in the context of a late time expansion for 
the scalar mode functions, rather than this early time expansion for the
tensor mode functions) that simply using the first order correction of the
mode function to infer the power spectrum does not give a very accurate
result \cite{Dvorkin:2009ne}. From (\ref{badcomp}) we can see that it also
gives the misleading impression that the correction to $M(t,k)$ is nonlocal,
whereas one can see from expression (\ref{ubeta}) that part of the second 
order correction exactly cancels this, leaving a purely local correction to
$M(t,k)$. 

\subsection{Factoring out the constant $\epsilon$ part}

Reflection on the early time expansion (\ref{Mexp}-\ref{betaM}) leads
to the following form for the terms which include no derivatives of 
$\epsilon(t)$,
\begin{eqnarray}
M_0(t,k) & = & \frac1{2 k a^2(t)} \Biggl\{ 1 + \sum_{n=1}^{\infty}
f_n\Bigl( \epsilon(t)\Bigr) \Bigl[ \frac{H(t) a(t)}{k}\Bigr]^{2n} 
\Biggr\} \; , \\
f_n(\epsilon) & \equiv & \frac{(2n \!-\! 1)!!}{(2 n)!!} \Bigl[(n\!+\!1)
\!-\! n\epsilon\Bigr] \Bigl[n \!-\! (n \!-\! 1)\epsilon \Bigr] \cdots
\Bigl[3 \!-\! 2 \epsilon\Bigr] \Bigl[2 \!-\! \epsilon\Bigr] \nonumber \\
& & \hspace{-.3cm} \times \! \Bigl[(n\!-\!1)\epsilon \!-\! (n\!-\!2)
\Bigr] \Bigl[(n\!-\!2)\epsilon \!-\! (n\!-\!3)\Bigr] \cdots 
\Bigl[\epsilon \!-\! 0 \Bigr] \Bigl[0 \!-\! (-1)\Bigr] \; . \qquad 
\end{eqnarray}
This is just $M_0(t,k) = \vert u_0(t,k)\vert^2$, where $u_0(t,k)$ is 
the constant $\epsilon$ solution (\ref{consteu}) evaluated at the 
instantaneous value of $\epsilon(t)$. The evolution of $\epsilon(t)$
is so slow in most cases that it makes sense to factor $M_0(t,k)$ out
of the result and derive an equation for the more sedate evolution of
the residual amplitude.

We begin by writing,
\begin{equation}
M(t,k) \equiv M_0(t,k) \times \Delta M(t,k) \qquad , \qquad
M_0(t,k) \equiv \vert u_0(t,k) \vert^2 \; . \label{ansatz}
\end{equation}
Differentiating (\ref{ansatz}) results in the relations,
\begin{eqnarray}
\dot{M} & = & \dot{M}_0 \times \Delta M + M_0 \times 
\Delta \dot{M} \; , \label{relation1} \\
\ddot{M} & = & \ddot{M}_0 \times \Delta M + 2 \dot{M}_0 \times
\Delta \dot{M} + M_0 \Delta \ddot{M} \; , \label{relation2} \\
\frac{\dot{M}^2}{2 M} & = & \frac{\dot{M}_0^2}{2 M_0} \times 
\Delta M + \dot{M}_0 \times \Delta \dot{M} + M_0 \times
\frac{\Delta \dot{M}^2}{2 \Delta M} \; , \label{relation3} \\
\frac1{2 a^6 M} & = & \frac1{2 a^6 M_0} \times \Delta M +
\frac1{2 a^6 M_0} \times \Bigl[-\Delta M + \frac1{\Delta M} 
\Bigr] \; . \label{relation4}
\end{eqnarray}
Substituting relations (\ref{relation1}-\ref{relation4}) into
(\ref{Meqn}) and dividing by $M(t,k)$ gives,
\begin{eqnarray}
\lefteqn{ \frac{\Delta \ddot{M}}{\Delta M} + \Bigl[ 3 H + 
\frac{\dot{M}_0}{M_0} \Bigr] \frac{\Delta \dot{M}}{\Delta M}
- \frac12 \Bigl(\frac{ \Delta \dot{M}}{\Delta M}\Bigr)^2 
+ \frac1{2 a^6 M_0^2} \Bigl[1 \!-\! \frac1{\Delta M^2} \Bigr] } 
\nonumber \\
& & \hspace{1.5cm} = -\frac{\ddot{M}_0}{M_0} - 3 H 
\frac{\dot{M}_0}{M_0} - \frac{2 k^2}{a^2} + \frac12 \Bigl( 
\frac{\dot{M_0}}{M_0}\Bigr)^2 + \frac1{2 a^6 M_0^2} 
\equiv S(t,k) \; . \qquad \label{DMeqn1}
\end{eqnarray}
This is an evolution equation for $\Delta M(t,k)$, which is
driven by a source $S(t,k)$. From (\ref{Mexp}-\ref{betaM}) we 
see that the early time expansion of $\Delta M(t,k)$ is,
\begin{equation}
\Delta M(t,k) = 1 + \Bigl[ \frac{9 \dot{\epsilon}}{8 H} - 
\frac{3  \epsilon \dot{\epsilon}}{4 H} + 
\frac{\ddot{\epsilon}}{8 H^2} \Bigr] \Bigl( \frac{ a H}{k}
\Bigr)^4 + O\Bigl( \frac{a^6 H^6}{k^6}\Bigr) \; . \label{earlyDM}
\end{equation}

\subsection{Simplifications}

Because $M_0(t,k)$ is an exact solution for constant $\epsilon(t)$,
it must be that the source $S(t,k)$ is proportional to $\dot{\epsilon}$ 
and $\ddot{\epsilon}$. This is not obvious from expression (\ref{DMeqn1})
because of the complicated way $M_0(t,k)$ depends upon time explicitly
through $a(t)$ and implicitly through $z(t,k)$ and $\nu(t)$,
\begin{equation}
M_0(t,k) = \frac{\frac{\pi z}{2} \!\times\! \vert 
H^{(1)}_{\nu}(z)\vert^2}{2 k a^2(t)} \quad , \quad z(t,k) \equiv 
\frac{k}{(1 \!-\! \epsilon) H a} \quad , \quad \nu(t) \equiv \frac12 
\!+\! \frac1{1 \!-\! \epsilon} \; . \label{M0def} 
\end{equation}
In Appendix A we make the following simplifications:
\begin{enumerate}
\item{Define the $z$ and $\nu$ dependent part of $M_0$ as $\sigma(z,\nu)
\equiv \ln[2 k a^2(t) \times M_0(t,k)]$;}
\item{Use the chain rule to express time derivatives of $M_0(t,k)$ as
$z$ and $\nu$ derivatives of $\sigma(z,\nu)$ multiplied by time derivatives
of $z(t,k)$ and $\nu(t)$;}
\item{Use Bessel's equation to eliminate the second $z$ derivative;}
\item{Change variables in $\sigma(z,\nu)$ from $z$ to $\zeta \equiv \ln(z)$ 
and from $\nu \equiv \frac12 + \Delta \nu$ to $\xi \equiv \ln[\Delta \nu]$;}
\item{Change the evolution variable from co-moving time $t$ to the number of 
e-foldings $N \equiv \ln[a(t)/a_i]$; and}
\item{Express $\Delta M(t,k)$ in terms of a new dependent variable 
$h(t,k)$ as $\Delta M(t,k) \equiv \exp[-\frac12 h(t,k)]$.}
\end{enumerate} 
When all of these things are done equation (\ref{DMeqn1}) takes the form,
\begin{eqnarray}
\lefteqn{ \partial^2_N h - \Bigl[\frac12 \partial_N h\Bigr]^2 + 
\Bigl[1 \!-\! \epsilon \!+\! \partial_N \sigma\Bigr] \partial_N h 
+ \Bigl[2 (1 \!-\! \epsilon) e^{\zeta - \sigma}\Bigr]^2 
\Bigl[e^{h} \!-\! 1\Bigr] } \nonumber \\
& & \hspace{-.5cm} = 2 \Bigl[ \frac{\partial_N^2 
\epsilon}{1 \!-\! \epsilon} \!+\! 2 \Bigl( \frac{\partial_N
\epsilon}{1 \!-\! \epsilon}\Bigr)^2\Bigr] \frac{\partial \sigma}{
\partial \zeta} + 2 \Bigl[ \partial_N \epsilon \!+\! 
\frac{\partial_N^2 \epsilon}{1 \!-\! \epsilon} \!+\! 2 \Bigl( 
\frac{\partial_N \epsilon}{1 \!-\! \epsilon}\Bigr)^2\Bigr] 
\frac{\partial \sigma}{\partial \xi} \nonumber \\
& & \hspace{-.3cm} + 4 \Bigl[ -\partial_{N}
\epsilon \!+\! \Bigl( \frac{\partial_N \epsilon}{1 \!-\! \epsilon}
\Bigr)^2 \Bigr] \Bigl[ \frac{\partial^2 \sigma}{\partial
\zeta \partial \xi} \!+\! \frac12 \frac{\partial \sigma}{\partial 
\zeta} \frac{\partial \sigma}{\partial \xi} \Bigr] 
+ 2 \Bigl( \frac{\partial_N \epsilon}{1 \!-\! \epsilon}\Bigr)^2 
\Bigl[\frac{\partial^2 \sigma}{\partial \xi^2} \!-\! \frac{\partial 
\sigma}{\partial \xi} \!+\! \frac12 \Bigl( \frac{\partial \sigma}{
\partial \xi}\Bigr)^2 \Bigr] \nonumber \\
& & \hspace{3.5cm} + 4 \Bigl[ -2\partial_N \epsilon \!+\! \Bigl( 
\frac{\partial_N \epsilon}{1 \!-\! \epsilon} \Bigr)^2 \Bigr] 
\Bigl[ \frac{(2 \!-\! \epsilon)}{(1 \!-\! \epsilon)^2} + e^{2 \zeta} 
(e^{-2\sigma} \!-\! 1)\Bigr] \; . \qquad \label{finaleqn}
\end{eqnarray}
If desired, the derivative of $\sigma$ with respect to $N$ on the 
first line of (\ref{finaleqn}) can be expressed like the terms on 
the right hand side of the equation,
\begin{equation}
\partial_N \sigma = \Bigl[-(1 \!-\! \epsilon) + \frac{\partial_N
\epsilon}{1 \!-\! \epsilon} \Bigr] \frac{\partial \sigma}{\partial
\zeta} + \frac{\partial_N \epsilon}{1 \!-\! \epsilon} 
\frac{\partial \sigma}{\partial \xi} \; .
\end{equation}

If $N_k$ represents the e-folding at which first horizon crossing 
occurs then one can express the scale factor in terms of $\Delta N
\equiv N - N_k$,
\begin{equation}
a = a_i e^N = a_i e^{N_k} \!\times\! e^{\Delta N} = 
\frac{k e^{\Delta N}}{H(t_k)} \; .
\end{equation}
Hence we have,
\begin{equation}
M(t,k) = \frac{e^{\sigma - \frac12 h}}{2 k a^2} =
\frac{H^2(t_k) C(\epsilon_k)}{2 k^3} \!\times\!
\exp\Bigl[ \sigma \!-\! \ln[C(\epsilon_k)] \!-\! 2 \Delta N \!-\!
\frac12 h\Bigr] \; , \label{Mfromh}
\end{equation}
where (\ref{epscor}) gives $C(\epsilon)$. The correction to the constant 
$\epsilon$ prediction we are seeking is the late time limit of the 
exponential factor in expression (\ref{Mfromh}).
 
\subsection{Asymptotic analysis}

In using equation (\ref{finaleqn}) it is important to understand its
limiting forms for early times ($k \gg H a$) and for late times ($k \ll
H a$). At early times $z(t,k)$ is large and $h(t,k)$ is small. In 
Appendix B we expand each of the factors of equation (\ref{finaleqn})
to show that its early time limiting form is,
\begin{eqnarray}
\lefteqn{\partial_N^2 h + (1 \!-\! \epsilon) \partial_N h + 
4 (1 \!-\! \epsilon)^2 z^2 h + O\Bigl(z^0 \times h\Bigr) } 
\nonumber \\
& & \hspace{4.3cm} = -\Bigl[2 (\nu \!+\! 3) \partial_N \epsilon + 
\frac{\partial_N^2 \epsilon}{1 \!-\! \epsilon}\Bigr] 
\frac{(\nu \!-\! \frac12)}{z^2} + O\Bigl(\frac1{z^4}\Bigr) \; .
\qquad \label{early}
\end{eqnarray}

Equation (\ref{early}) represents a damped, driven oscillator with,
\begin{eqnarray}
{\rm Friction\ Force} & \Longrightarrow & -(1 \!-\! \epsilon) \times 
\partial_N h \; , \\
{\rm Restoring\ Force} & \Longrightarrow & -4 (1 \!-\! \epsilon)^2 
z^2 \times h \; , \label{restore} \\
{\rm Driving\ Force} & \Longrightarrow & -\Bigl[ 2(\nu \!+\! 3) 
\partial_N \epsilon + \frac{\partial_N^2 \epsilon}{1 \!-\! 
\epsilon}\Bigr] \frac{(\nu \!-\! \frac12)}{z^2} \; . \qquad
\label{drive}
\end{eqnarray}
The restoring force (\ref{restore}) pushes $h(t,k)$ down to zero if 
it ever gets displaced. The driving force (\ref{drive}) does push 
$h(t,k)$ away from zero, but its coefficient falls like $1/z^2$
whereas the restoring force grows like $z^2$. The ``time'' (that is, $N$)
derivatives are irrelevant at leading order in $z$, so the result
in this regime is just the local ``tracking relation'' we noted in
expression (\ref{earlyDM}),
\begin{eqnarray}
h(t,k) & = & -\Bigl[ 2(\nu \!+\! 3) \partial_N \epsilon + 
\frac{\partial_N^2 \epsilon}{1 \!-\! \epsilon}\Bigr] \frac{(\nu \!-\! 
\frac12)}{4 (1 \!-\! \epsilon)^2 z^4} + O\Bigl( \frac1{z^6}\Bigr) 
\; , \\
& = & -\frac14 \Bigl[ (9 \!-\! 7 \epsilon) \partial_N \epsilon +
\partial_N^2 \epsilon \Bigr] \Bigl( \frac{H a}{k}\Bigr)^4 + 
O\Bigl(\frac{H^6 a^6}{k^6}\Bigr) \; . \label{initial}
\end{eqnarray}
This explains why the early time expansion is local to all orders.
It also explains the striking property of Figure~\ref{jump} that 
an instantaneous jump in $\epsilon(t)$ --- which makes the source 
{\it diverge} --- has negligible effect until just a few e-foldings 
before first horizon crossing. One consequence is that we may as well
begin numerical evolution at $N = N_k - 7$ using expansion 
(\ref{initial}) to determine the initial values of $h(t,k)$ and 
$\partial_N h(t,k)$.

At late times $z(t,k)$ is small but $h(t,k)$ can grow to reach 
significant values. In Appendix C we expand the various factors of 
(\ref{finaleqn}) to show that the late time limiting form is,
\begin{eqnarray}
\lefteqn{ \partial_N^2 h - \Bigl[ \frac12 \partial_N h\Bigr]^2 +
\Bigl[2 \partial_N \epsilon \Delta \nu^2 F \!+\! 3 \!-\! \epsilon
\Bigr] \partial_N h = 4 \partial_N \epsilon \Delta \nu \Bigl[(2 \Delta
\nu \!+\! 1) F +1 \Bigr] } \nonumber \\
& & \hspace{-.7cm} + 4 \Bigl(\frac{\partial_N^2 \epsilon}{1 \!-\!
\epsilon}\Bigr) \Delta \nu F \!+\! 4 \Bigl( \frac{ \partial_N \epsilon}{1 
\!-\! \epsilon} \Bigr)^2 \!\! \Delta \nu \Bigl[ \Delta \nu F^2 \!+\! 2 F
\!-\! 1 \!+\! \Delta \nu \psi'\Bigl(\frac12 \!+\! \Delta \nu\Bigr)\!\Bigr] 
\!+\! O(z^2) . \qquad \label{lateqn}
\end{eqnarray}
Here $F(t,k)$ stands for the quantity,
\begin{equation}
F \equiv -1 \!-\! \ln\Bigl(\frac{z}{2}\Bigr) \!+\! 
\psi\Bigl( \frac12 \!+\! \Delta \nu\Bigr) = \Delta N \!-\! 1
\!+\! \ln\Bigl[ \frac{2(1 \!-\! \epsilon) H}{H(t_k)}\Bigr] \!+\! 
\psi\Bigl( \frac12 \!+\! \frac1{1 \!-\! \epsilon}\Bigr) \; . 
\label{Fdef}
\end{equation}

The late time equation (\ref{lateqn}) implies,
\begin{eqnarray}
\partial_N h & = & 4 \partial_N \epsilon \Delta \nu^2 F + 
O(z^2) \; , \label{latedNh} \\
\partial_N^2 h & = & 4 \Bigl[ \frac{\partial_N^2 \epsilon}{1
\!-\! \epsilon} + 2 \Bigl(\frac{\partial_N \epsilon}{1 \!-\!
\epsilon}\Bigr)^2 \Bigr] \Delta \nu F + 4 \partial_N \epsilon
\Delta \nu \nonumber \\
& & \hspace{2.5cm} + 4 \Bigl( \frac{\partial_N \epsilon}{1 \!-\!
\epsilon}\Bigr)^2 \Delta \nu \Bigl[-1 \!+\! \Delta \nu 
\psi'\Bigl(\frac12 \!+\! \Delta \nu\Bigr)\Bigr] + O(z^2) \; . 
\qquad 
\end{eqnarray}
Hence the asymptotic form of $h(t,k)$ at late times is,
\begin{equation}
h(t,k) = 4 \Delta \nu \epsilon \Delta N \!+ 4 \Delta \nu \ln\Bigl[
\frac{H}{H(t_k)}\Bigr] + 2 \ln\Bigl[ \frac{C(\epsilon)}{C(\epsilon_k)}
\Bigr] - 2 \ln\Bigl[ \mathcal{C}(k)\Bigr] + O(z^2) \; . \label{final}
\end{equation}
Comparison with (\ref{Mfromh}) reveals the unknown constant 
$\mathcal{C}(k)$ as the correction factor we seek to the constant 
$\epsilon$ prediction for the tensor power spectrum.

\subsection{An analytic approximation for $\Delta M(t,k) = 
e^{-\frac12 h(t,k)}$}

\begin{figure}[ht]
\includegraphics[width=10.0cm,height=8.0cm]{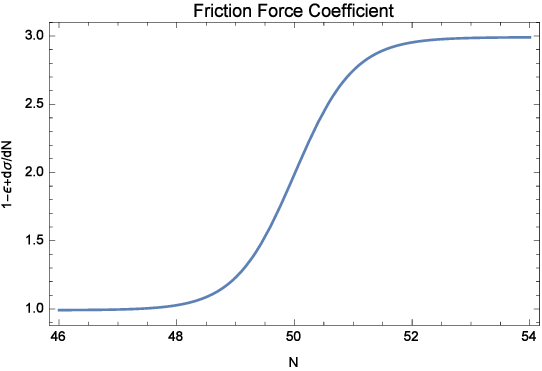}
\caption{Graph of $(1 \!-\! \epsilon \!+\! \partial_N \sigma)$ as a 
function of $N$, assuming $N_k = 50$ and $\epsilon(N) = \frac1{200 - 
2N}$, which corresponds to $V(\varphi) \propto \varphi^2$.}
\label{friction}
\end{figure}

\begin{figure}[ht]
\includegraphics[width=6.0cm,height=5.0cm]{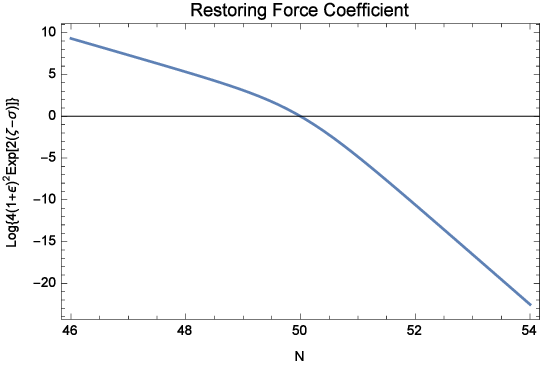} \hspace{1cm}
\includegraphics[width=6.0cm,height=5.0cm]{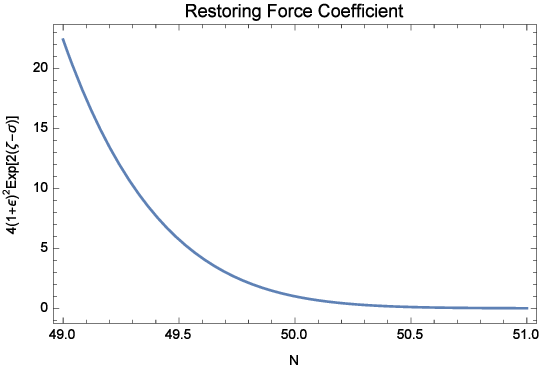}
\caption{Log (left) and linear (right) plots of $[2(1\!-\!\epsilon)
e^{\zeta-\sigma}]^2$ as a function of $N$, assuming $N_k = 50$ and 
$\epsilon(N) = \frac1{200 - 2N}$, which corresponds to $V(\varphi) 
\propto \varphi^2$.}
\label{forceAB}
\end{figure}

The behaviors we noted in the previous section are generic, and
they imply that we only need to bridge a small range of e-foldings
around first horizon crossing $N_k$ to carry the early form 
(\ref{initial}) into the late form (\ref{final}). In this region 
$h(t,k)$ is small and we can linearize equation (\ref{finaleqn}),
\begin{equation}
\partial_N^2 h + \Bigl[1 \!-\! \epsilon \!+\! \partial_N \sigma\Bigr]
\partial_N h + \Bigl[2 (1\!-\!\epsilon) e^{\zeta -\sigma}\Bigr]^2 h
\approx \mathcal{S}(N,N_k) \; , \label{lineqn}
\end{equation}
where $\mathcal{S}(N,N_k)$ is the full source term on the right hand 
side of (\ref{finaleqn}). Just like the early time form (\ref{early}),
equation (\ref{lineqn}) is a damped, driven harmonic oscillator.
Figure~\ref{friction} shows the friction term for the $V = \frac12 m^2 
\varphi^2$ model. Figure~\ref{forceAB} gives log and linear plots of 
the restoring force for the same model.

It is easy to develop a Green's function solution to (\ref{lineqn}).
Note that the homogeneous equation takes the form,
\begin{equation}
\chi'' - \frac{\omega'}{\omega} \, \chi' + \omega^2 \chi = 0 \quad ,
\quad \chi' \equiv \partial_N \chi(N,N_k) \quad , \quad \omega'
\equiv \partial_N \omega(N,N_k) \; , \label{homogeneous}
\end{equation}
where the frequency is,
\begin{equation}
\omega(N,N_k) \equiv 2 (1 \!-\! \epsilon) e^{\zeta - \sigma} \; .
\label{omega}
\end{equation}
The two linearly independent solutions of (\ref{homogeneous}) can
be expressed in terms of the integral of $\omega(N,N_k)$,
\begin{equation}
\chi_{\pm}(N,N_k) = \exp\Bigl[ \pm i \!\! \int_0^N \!\!\!\!\! dn \, 
\omega(n,N_k)\Bigr] \quad \Longrightarrow \quad
\chi_+' \chi_- - \chi_+ \chi_-' = 2i \omega \; . \label{twosols}
\end{equation}
Hence the retarded Green's function we seek is,
\begin{equation}
G(N;N') = \frac{\theta(N \!-\! N')}{\omega(N',N_k)} \, \sin\Bigl[
\int_{N'}^{N} \!\!\!\!\! dn \, \omega(n,N_k)\Bigr] \; . 
\label{Gfunction}
\end{equation}
And the Green's function solution to (\ref{lineqn}) is,
\begin{equation}
h(t,k) = \int_{0}^{N} \!\!\!\!\! dn \, \sin\Bigl[\int_{n}^{N} 
\!\!\!\!\! dn' \, \omega(n',N_k)\Bigr] \frac{\mathcal{S}(n,N_k)}{
\omega(n,N_k)} \; . \label{GFsol1}
\end{equation}

The asymptotic expansion (\ref{initial}) is so accurate at early 
times that one may as well begin the evolution at some point near to 
first horizon crossing, say $N_1 = N_k - 7$. Then the Green's function 
solution takes the form,
\begin{eqnarray}
\lefteqn{h(t,k) = \cos\Bigl[\int_{N_1}^{N} \!\!\!\!\! dn \, 
\omega(n,N_k)\Bigr] \!\times\! h(t_1,k) + \sin\Bigl[\int_{N_1}^{N} 
\!\!\!\!\! dn \, \omega(n,N_k)\Bigr] \!\times\! 
\frac{\partial_N h(t_1,N_k)}{\omega(N_1,N_k)} } \nonumber \\
& & \hspace{5cm} + \int_{N_1}^{N} \!\!\!\!\! dn \, \sin\Bigl[\int_{n}^{N} 
\!\!\!\!\! dn' \, \omega(n',N_k)\Bigr] \frac{\mathcal{S}(n,N_k)}{
\omega(n,N_k)} \; . \qquad \label{GFsol2}
\end{eqnarray}
The initial values $h(t_1,k)$ and $\partial_N h(t_1,k)$ can either be 
computed from (\ref{initial}) or simply approximated as zero.

Whether one uses expression (\ref{GFsol1}) or (\ref{GFsol2}), the goal 
is to evolve it to some point safely after first horizon crossing, say
$N_2 = N_k + 7$. Then the nonlocal correction factor $\mathcal{C}(k)$
can be estimated by ignoring the order $z^2$ terms in expression 
(\ref{final}),
\begin{equation}
\mathcal{C}(k) \approx \exp\Biggl[2 \Delta \nu_2 \epsilon_2 \Delta N_2
\!+\! 2\Delta \nu_2 \ln\Bigl[ \frac{H(t_2)}{H(t_k)}\Bigr] \!+\! 
\ln\Bigl[ \frac{C(\epsilon_2)}{C(\epsilon_k)}\Bigr] \!-\! \frac12
h(t_2,k)\Biggr] . \label{bestest}
\end{equation}
Expression (\ref{bestest}) is radically different from other numerical 
schemes for computing the tensor power spectrum in that it gives an 
approximate but closed form expression for arbitrary first slow roll 
parameter $\epsilon(N)$. One consequence is that we can use the 
transformation (\ref{MtoN}) to immediately read off the analogous 
correction to the constant $\epsilon$ prediction (\ref{scalarconsteps})
for the scalar power power spectrum. Expression (\ref{bestest}) is also 
the best way of deconvolving features in the power spectrum 
\cite{Chung:1999ve,Elgaroy:2003hp} to reconstruct the geometrical 
conditions which produced them.

\section{Numerical Analyses}\label{numerical}

The purpose of this section is to support various conclusions using 
numerical solutions of our full equation (\ref{finaleqn}) for $h(t,k)$.
Recall that the full amplitude is given by $M(t,k) = M_0(t,k) \times
\exp[-\frac12 h(t,k)]$, where $M_0(t,k)$ is the known constant 
$\epsilon$ solution (\ref{M0def}). Recall also that the ultimate 
observable is the correction factor $\mathcal{C}(k)$ --- inferred from
$h(t,k)$ using expression (\ref{final}) --- to the constant $\epsilon$ 
approximation (\ref{tensorconsteps}) for the tensor power spectrum.

\subsection{$\mathcal{C}(k)\! - \!1$ is small for smooth models}

\begin{figure}[ht]
\includegraphics[width=10.0cm,height=8.0cm]{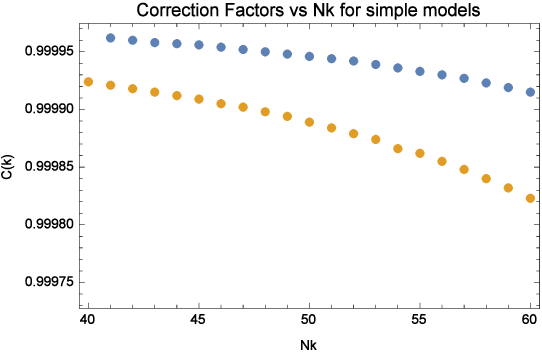}
\vspace{0cm}
\caption{Correction factors $\mathcal{C}(k)$ to the constant $\epsilon$ 
approximation for $\epsilon(N) = [200 \!-\! 2N]^{-1}$ (blue), corresponding
$V(\varphi) \propto \varphi^2$, and for $\epsilon(N) = [100 \!-\! N]^{-1}$
(yellow), corresponding to $V(\varphi) \propto \varphi^4$.}
\label{Simple}
\end{figure}

It has long been obvious the constant $\epsilon$ approximation
(\ref{scalarconsteps}-\ref{tensorconsteps}) are wonderfully accurate 
for models in which $\epsilon$ is small and varies smoothly near first
horizon crossing \cite{Wang:1997cw}. Figure~\ref{Simple} confirms this
for two simple monomial potentials,
\begin{eqnarray}
V(\varphi) \propto \varphi^2 & \Longrightarrow &
\epsilon(N) = \frac1{200 \!-\! 2N} \; , \\
V(\varphi) \propto \varphi^4 & \Longrightarrow &
\epsilon(N) = \frac1{100 \!-\! N} \; .
\end{eqnarray}
Figure~\ref{Simple} also answers the first of the questions posed at
the end of the Introduction: it seems that the constant $\epsilon$ 
approximation is most accurate for $\epsilon$ near to $\epsilon_k$.
One can see this by comparing the value of the correction factor 
$C(\epsilon)$, defined in (\ref{epscor}), with the nonlocal correction 
factor $\mathcal{C}(k)$ shown in Figure~\ref{Simple}, over the 20 
e-foldings of first horizon crossing ($40 < N_k < 60$) depicted,
\begin{eqnarray}
V(\varphi) \propto \varphi^2 & \Longrightarrow & \Biggl\{
{0.99546 < C(\epsilon_k) < 0.99317 \atop 0.99996 < \mathcal{C}(k) < 0.99991} 
\Biggr\} \; , \\
V(\varphi) \propto \varphi^4 & \Longrightarrow & \Biggl\{
{0.99084 < C(\epsilon_k) < 0.98619 \atop 0.99992 < \mathcal{C}(k) < 0.99982}
\Biggr\} \; .
\end{eqnarray}
There is about 50 times more variation in $C(\epsilon_k)$ than in 
$\mathcal{C}(k)$, limiting the potential improvement to a positive offset of 
about $\Delta N \approx \frac{20}{50} = 0.4$. Because other models show 
$\mathcal{C}(k) > 1$ there is actually no preference for shifting the point
at which $\epsilon$ is evaluated.
 
\vspace{0cm} 

\subsection{$\mathcal{C}(k) \!-\!1$ significant for changes near horizon crossing}

\begin{figure}[ht]
\includegraphics[width=6.0cm,height=4.0cm]{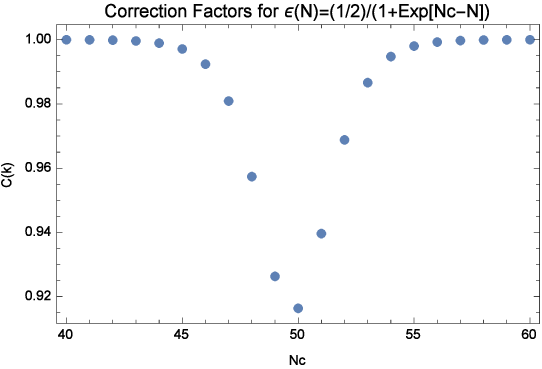}\hspace{1cm}
\includegraphics[width=6.0cm,height=4.0cm]{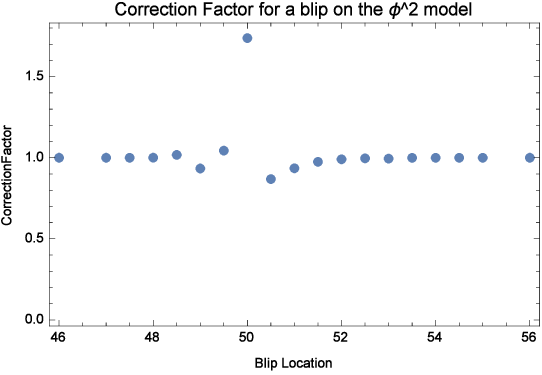}
\vspace{0cm}
\caption{Correction factors $\mathcal{C}(k)$ to the constant $\epsilon$ 
approximation for two models with smooth transitions centered at an
arbitrary point $N_c$. The first model has $\epsilon(N) = \frac1{2 [1 + 
e^{N_c - N}]}$, corresponding to the left hand graph. The right hand 
graph corresponds to a $V(\varphi) \propto \varphi^2$ which experiences a 
Gaussian ``blip'' defined by (\ref{blipmodel}). In each case horizon 
crossing is fixed at $N_k = 50$ and the graph shows how $\mathcal{C}(k)$
changes as $N_c$ varies.}
\label{Logistic}
\end{figure}

It has also long been understood that the constant $\epsilon$ formulae 
require significant corrections when $\epsilon(N)$ suffers large variation
within several e-foldings of first horizon crossing \cite{Starobinsky:1992ts,
Mukhanov:1991rp}. We already saw this in the exact results depricted in
Figure~\ref{jump} for an instantaneous jump in $\epsilon$. Figure~\ref{Logistic} 
makes the same point for two smooth transitions. The left hand graph shows the
effect on $\mathcal{C}(k)$ of a smooth transition from $\epsilon = 0$ to 
$\epsilon = \frac12$ via a logistic function centered at a critical value $N_c$,
\begin{equation}
\epsilon(N) = \frac{0.5}{1 + e^{N_c - N}} \; .
\end{equation}
The right hand graph shows $\mathcal{C}(k)$ for a $V(\varphi) \propto 
\varphi^2$ model which experiences a Gaussian bump, centered at $N_c$, which
actually induces a brief deceleration,
\begin{equation}
\epsilon(N) = \frac1{200 \!-\! 2 N} + \exp\Bigl[-10 (N \!-\! N_c)^2 \Bigr] 
\; . \label{blipmodel}
\end{equation}
This is one of the models for which $\mathcal{C}(k)$ is larger than one. 
\vspace{0cm} 

\subsection{Eqn. (\ref{GFsol2}) is quite accurate near horizon crossing}

\begin{figure}[ht]
\includegraphics[width=6.0cm,height=4.0cm]{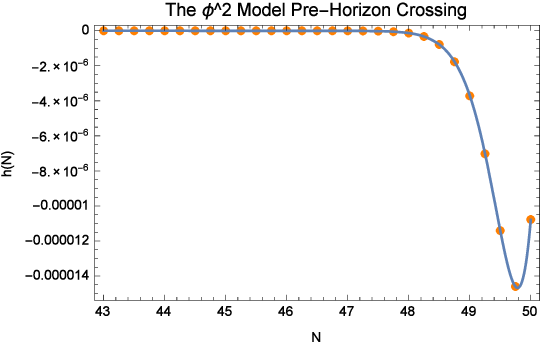}\hspace{1cm}
\includegraphics[width=6.0cm,height=4.0cm]{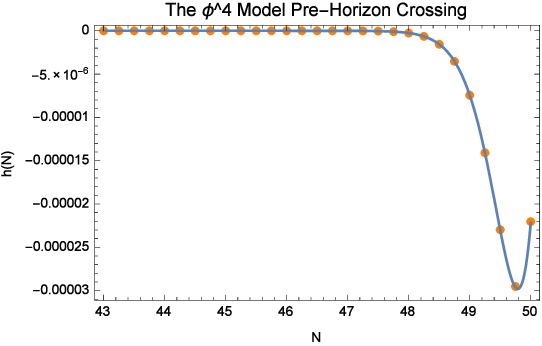}
\vspace{0cm}
\caption{The pre-horizon crossing regime of $h(t,k)$ for two simple models.
The left hand graph shows $\epsilon(N) = [200 \!-\! 2 N]^{-1}$, corresponding 
to $V(\varphi) \propto \varphi^2$, and the right hand graph shows $\epsilon(N) 
= [100 \!-\! N]^{-1}$, corresponding to $V(\varphi) \propto \varphi^4$. In
each case the continuous blue line represents numerical evolution of the full
nonlinear equation (\ref{finaleqn}) and the yellow dots give the analytic
approximation (\ref{GFsol2}).}
\label{Linearmodels}
\end{figure}

The previous two points were known before in general terms. Our contributions
in this paper are:
\begin{enumerate}
\item{An analytic quantification --- through the asymptotic expansions
(\ref{initial}) and (\ref{final}) --- of when to expect significant corrections
to the constant $\epsilon$ approximation; and}
\item{An analytic approximation (\ref{GFsol2}) of the function $h(t,k)$ which
gives us the nonlocal correction factor through expression (\ref{bestest}).}
\end{enumerate}
Figure~\ref{Linearmodels} show just how accurate our approximation is in the 
period before first horizon crossing. It even catches the turning points at 
$N \sim 49.8$.
\vspace{0cm} 
\vspace{0cm} 

\vspace{0cm} 
\begin{figure}[ht]
\includegraphics[width=4.0cm,height=3.0cm]{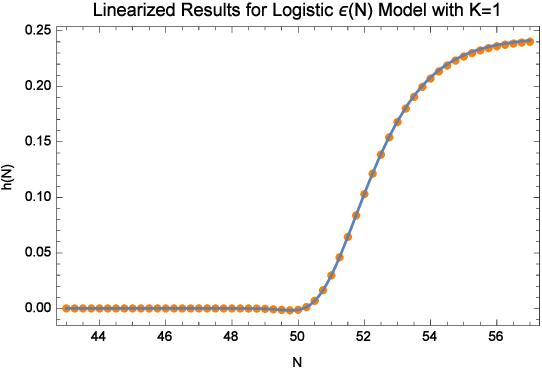}
\hspace{.5cm}
\includegraphics[width=4.0cm,height=3.0cm]{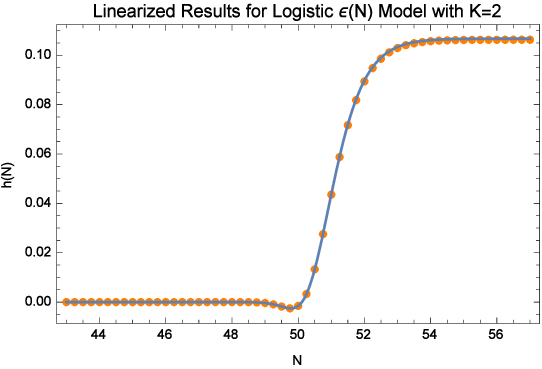}
\hspace{.5cm}
\includegraphics[width=4.0cm,height=3.0cm]{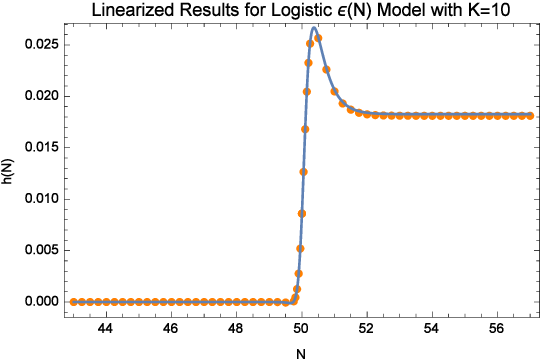}
\vspace{0cm}
\caption{The pre and post horizon crossing regimes for $\epsilon(N) = 0.005 +
0.095 \!\times\! (1 \!+ \exp[-K \!\times\! (N \!-\! N_k)])^{-1}$, for $K=1$
(left), $K=2$ (center) and $K=10$ (right). Each of these models interpolates
between $\epsilon = 0.005$ at early times to $\epsilon = 0.100$ at late times,
with $\epsilon_k =0.0 525$. The continuous blue line represents numerical 
evolution of the full nonlinear equation (\ref{finaleqn}) and the yellow 
dots give the analytic approximation (\ref{GFsol2}).}
\label{LinearizedLogisticEpsilon}
\end{figure}

Our analytic approximation (\ref{GFsol2}) continues to be very accurate 
after first horizon crossing for models in which there is no significant 
evolution of  $\epsilon(t)$ at late times. Figure~\ref{LinearizedLogisticEpsilon} 
illustrates this by showing $h(t,k)$ versus $N$ for a class of models in
which $\epsilon$ makes a transition (centered about horizon crossing of
$N_k = 50$) from an early value of $\epsilon = \frac1{200}$ to a late value 
of $\epsilon = \frac1{10}$ through a logistic function with steepness 
parameter $K = 1, 2, 10$,
\begin{equation}
\epsilon(N) = \frac1{200} + \frac{\frac{19}{200}}{1 \!+\! \exp[-K (N \!-\!
N_k)]} \; .
\end{equation}
In each case the horizon crossing value is $\epsilon_k = \frac{41}{400}$. 

Note from Figure~\ref{LinearizedLogisticEpsilon} that the final value of 
$h(t,k)$ is largest when the transition is most gradual. Because the
full amplitude is $M(t,k) = M_0(t,k) \!\times\! \exp[-\frac12 h(t,k)]$,
one might expect that $M(t,k)$ therefore freezes in to a smaller amplitude 
for a more gradual transition. In fact the reverse is true because only
the value of $\epsilon$ near horizon crossing is relevant, so making
$\epsilon$ stay small longer causes the freeze-in amplitude to be larger. 
This is evident from the nonlocal correction factors $\mathcal{C}(k)$ for 
the three cases,
\begin{eqnarray}
K=1 & \Longrightarrow & \mathcal{C}(k) = 0.993556 \qquad (0.994121) 
\; , \label{mathC1} \\
K=2 & \Longrightarrow & \mathcal{C}(k) = 0.989201 \qquad (0.989418) 
\; , \label{mathC2} \\
K=10 & \Longrightarrow & \mathcal{C}(k) = 0.975152 \qquad (0.975230) 
\; . \label{mathC10}
\end{eqnarray}
(The parenthesized values are for the linearized approximation, which
shows that it is indeed quite good.) We fixed the values of $H(t_k)$ and
$a(t_k)$ to be the same for each model, so these correction factors give 
the relative freeze-in amplitudes for $M(t,k)$. 

The much larger and opposite-sense effect which is evident in the 
asymptotic values of $h(t,k)$ of Figure~\ref{LinearizedLogisticEpsilon} 
is needed to compensate for the factor $M_0(t,k)$. Recall from section 
3.2 that if $\epsilon(t)$ becomes constant at $\epsilon_1$ for times 
$t > t_1$ then we can write,
\begin{equation}
\epsilon(t) = \epsilon_1 \qquad \Longrightarrow \qquad H(t) 
a^{\epsilon_1}(t) = H_1 a_1^{\epsilon_1} \; ,
\end{equation}
where $H_1 \equiv H(t_1)$ and $a_1 \equiv a(t_1)$. Each of the three
models has effectively reached this condition by 10 e-foldings after
first horizon crossing, but the values of $H_1$ and $a_1$ are smaller the 
steeper the transition. That affects the factor of $M_0(t,k)$, which 
approaches a constant given by (\ref{u0late}),
\begin{equation}
M_0(t,k) \longrightarrow \frac{H^2(t_k)}{2 k^3} \!\times\! C(\epsilon_1) 
\!\times\! \Biggl[ \frac{H_1 a_1^{\epsilon_1}}{H(t_k) a^{\epsilon_1}(t_k)}
\Biggr]^{\frac{2}{1-\epsilon_1}} \; . \label{logisticM0}
\end{equation}
The final factor of (\ref{logisticM0}) is significantly larger for
more gradual transitions, which is mostly cancelled by the larger 
asymptotic values of $h(t,k)$, to leave the small effect evident in  
the nonlocal correction factors (\ref{mathC1}-\ref{mathC10}).

\vspace{0cm} 
\begin{figure}[ht]
\includegraphics[width=6.0cm,height=4.0cm]{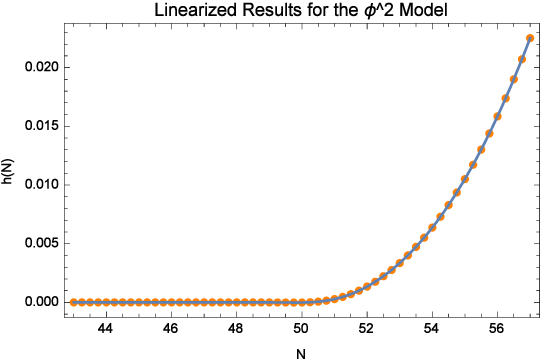}\hspace{1cm}
\includegraphics[width=6.0cm,height=4.0cm]{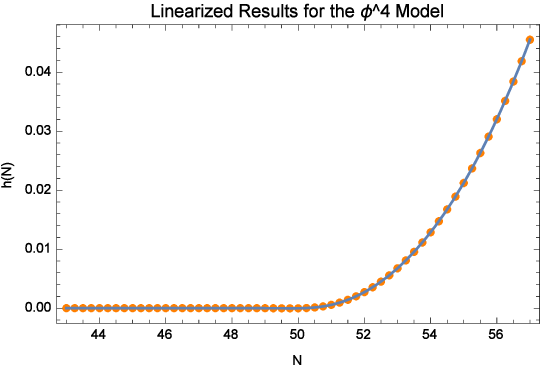}
\vspace{0cm}
\caption{The pre and post horizon crossing regimes of $h(t,k)$ for two simple 
models. The left hand graphs concerns $\epsilon(N) = [200 \!-\! 2 N]^{-1}$, 
corresponding to $V(\varphi) \propto \varphi^2$, and the right hand graph 
concerns $\epsilon(N) = [100 \!-\! N]^{-1}$, corresponding to $V(\varphi) 
\propto \varphi^4$. In each case the continuous blue line represents numerical 
evolution of the full nonlinear equation (\ref{finaleqn}) and the yellow dots 
give the analytic approximation (\ref{GFsol2}).}
\label{Linearmodelsv2}
\end{figure}

The same late time effect of $h(t,k)$ partially compensating for changes in 
$M_0(t,k)$ is evident from the results for $V(\varphi) \propto \varphi^2$ 
and $V(\varphi) \propto \varphi^4$ models displayed in Figure~\ref{Linearmodelsv2}. 
In this case $\epsilon(t)$ continues to evolve after first horizon crossing.
Considered as a function of $N$ we have $\partial_N \ln[H(N)] = -\epsilon(N)$,
so the asymptotic form (\ref{final}) can be re-expressed as,
\begin{equation}
h(t,k) = \frac{4}{1 \!-\! \epsilon(N)} \! \int_{N_k}^N \!\!\!\!\! dn \Delta n 
\epsilon'(n) + 2 \ln\Bigl[ \frac{C(\epsilon(N))}{C(\epsilon_k)}\Bigr] -
2 \ln\Bigl[ \mathcal{C}(k)\Bigr] + O\Bigl( e^{-2 \Delta N}\Bigr) \; , 
\label{newfinal}
\end{equation} 
where $\Delta n \equiv n - N_k$ and $\Delta N \equiv N - N_k$. Because 
$\epsilon$ typically grows slowly with $N$ (as it does for both of the models
in Figure~\ref{Linearmodelsv2}) the integral grows and dominates the slowly
falling logarithm of (\ref{newfinal}), so that $h(t,k)$ grows like $\Delta N^2$.
This growth is evident for both models in Figure~\ref{Linearmodelsv2}.

\subsection{Problems long after horizon crossing}

\begin{figure}[ht]
\includegraphics[width=6.0cm,height=4.0cm]{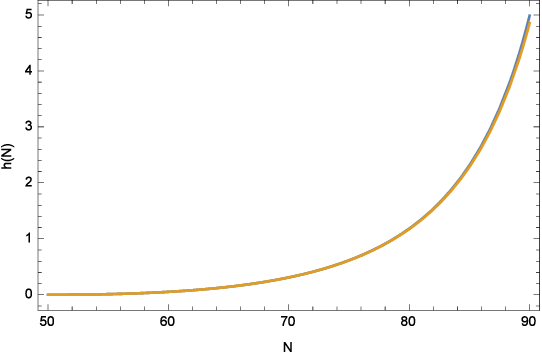}\hspace{1cm}
\includegraphics[width=6.0cm,height=4.0cm]{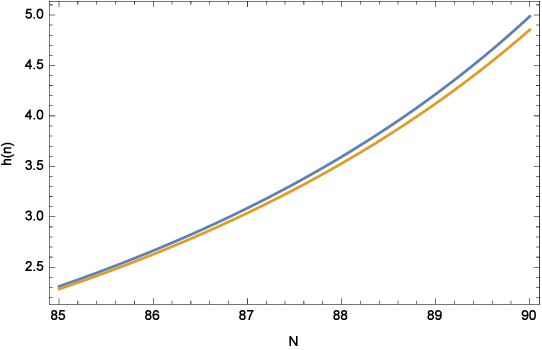}
\vspace{0cm}
\caption{Each graph shows the model with $\epsilon(N) = [200 \!-\! 2 N]^{-1}$,
corresponding to $V(\varphi) \propto \varphi^2$. Horizon crossing is at $N_k =
50$ and inflation ends at $N = 99.5$. In each case the blue line represents 
numerical evolution of the full nonlinear equation (\ref{finaleqn}) and the
yellow line gives our analytic approximation (\ref{GFsol2}).}
\label{Linearmodelsv3}
\end{figure}

Of course too much growth endangers the linearized approximation we made in 
passing from the full equation (\ref{finaleqn}) to (\ref{lineqn}). Recall
that this entails changing two terms,
\begin{eqnarray}
-\Bigl[ \frac12 \partial_N h\Bigr]^2 & \longrightarrow & 0 
\; , \label{approx1} \\
\exp\Bigl[ h(t,k)\Bigr] \!-\! 1 & \longrightarrow & h(t,k) 
\; . \label{approx2}
\end{eqnarray}
There is never any problem with (\ref{approx2}) because $h(t,k)$ is small
before first horizon crossing and the coefficient of this term is minuscule
after first horizon crossing. The problematic approximation is (\ref{approx1}),
although only in the region after first horizon crossing for models in which
$\epsilon$ evolves at very late times. One can see from expression (\ref{latedNh})
that two terms contribute to provide the factor of $F^2 \sim \Delta N^2$ 
(recall the definition (\ref{Fdef}) of $F$) which is evident in the late time 
evolution equation (\ref{lateqn}),
\begin{eqnarray}
-\Bigl[ \frac12 \partial_N h\Bigr]^2 & \longrightarrow & - \Bigl[2 \partial_N 
\epsilon \, \Delta \nu^2 F\Bigr]^2 \; , \label{factor1} \\
2 \partial_N \epsilon \, \Delta \nu^2 F \!\times\! \partial_N h 
& \longrightarrow & +2 \Bigl[2 \partial_N \epsilon \, \Delta \nu^2 F\Bigr]^2 
\; . \label{factor2}
\end{eqnarray}
These terms are enhanced by the factor of $F^2 \sim \Delta N^2$ but suppressed
by $(\partial_N \epsilon)^2$. In the full nonlinear equation (\ref{factor1}) 
cancels half of (\ref{factor2}), but this cancellation does not happen in the 
linearized equation because (\ref{factor1}) is not present. Hence we expect the 
very late time growth of the linearized approximation (\ref{GFsol2}) to be less
than what it is for the actual solution. This is barely evident in 
Figure~\ref{Linearmodelsv3} for the $V(\varphi) \propto \varphi^2$ model at 
very late times.

\vspace{0cm} 

The problem we have just described might seem serious but it is not. The
full amplitude $M(t,k) = M_0(t,k) \times \exp[-\frac12 h(t,k)]$ really does 
become constant shortly after first horizon crossing. The growth of $h(t,k)$ 
is only an artifact of the our having factored out by $M_0(t,k)$, which also 
grows for models in which $\epsilon$ increases at late times. Because the
problem has such a simple origin, there are two easy fixes:
\begin{enumerate}
\item{Either evaluate $\mathcal{C}(k)$ using expression (\ref{bestest}) at
some point $N_2$ before nonlinear effects become important; or}
\item{Subtract the right hand side of (\ref{factor1}) from the 
source $\mathcal{S}(N,N_k)$ in the Green's function solution.}
\end{enumerate}

\section{Epilogue}\label{epilogue}

The full scalar and tensor power spectra can be expressed in terms of
two amplitudes $N(t,k)$ and $M(t,k)$,
\begin{eqnarray}
\Delta^2_{\mathcal{R}}(k) & = & \frac{2 G k^3}{\pi} \!\times\! N\Bigl( 
t \!\gg\! t_k,k\Bigr) \!\times\! \Biggl\{1 + O(G H^2)\Biggr\} \; , \\
\Delta^2_{h}(k) & = & \frac{32 G k^3}{\pi} \!\times\! M\Bigl(t \!\gg\! 
t_k,k\Bigr) \!\times\! \Biggl\{1 + O(G H^2)\Biggr\} \; .
\end{eqnarray}
If the one loop corrections of order $G H^2 \ltwid 10^{-11}$ are ever 
to resolved we must have precise predictions for the two amplitudes.
Part of this problem entails finding a unique model for primordial 
inflation, which is beyond the scope of our present effort. We have
instead focussed on predicting how the amplitudes depend upon the
inflationary expansion history $a(t)$. Our analysis is based on earlier 
work in which nonlinear equations for the two amplitudes were derived
\cite{Romania:2011ez,Romania:2012tb}.

Because the transformation (\ref{MtoN}) carries $M(t,k)$ into $N(t,k)$, 
we worked with the simpler tensor amplitude. We express its late time
limiting form as,
\begin{equation}
M\Bigl(t \gg t_k,k\Bigr) = \frac{ H^2(t_k)}{2 k^3} \!\times\! 
C\Bigl(\epsilon(t_k)\Bigr) \!\times\! \mathcal{C}(k) \; ,
\end{equation}
where $C(\epsilon)$ was defined in expression (\ref{epscor}) and graphed
in Figure~\ref{epsilonfactor}. Our numerical studies show that this
factor really does need to present, and it is best to evaluate it at 
the time $t_k$ of first horizon crossing. The remaining factor 
$\mathcal{C}(k)$ represents nonlocal effects from the expansion
history before and after first horizon crossing. It has long been 
clear that this factor is close to unity for models in which
$\epsilon(t)$ is smooth near first horizon crossing, but $\mathcal{C}(k)$ 
can give significant corrections when there are large changes within a 
few e-foldings of first horizon crossing \cite{Starobinsky:1992ts,
Mukhanov:1991rp}.

Our key results (\ref{final}) and (\ref{bestest}) give, for the first 
time ever, a good analytic approximation for the nonlocal correction 
factor $\mathcal{C}(k)$. Our technique was to first get close to the 
exact solution by factoring out the known, constant $\epsilon$ solution 
$M_0(t,k)$,
\begin{equation}
M(t,k) = M_0(t,k) \!\times\! \exp\Bigl[-\frac12 h(t,k)\Bigr] \; .
\end{equation}
Of course this means that the evolution equation (\ref{finaleqn}) for 
$h(t,k)$ is driven by a source term which vanishes whenever $\epsilon(t)$ 
is constant. From the equation's asymptotic early time form (\ref{early}) 
we can see that that $h(t,k)$ behaves as a damped, driven harmonic 
oscillator. For more than a few e-foldings before first horizon crossing 
the restoring force (\ref{restore}) is so large that $h(t,k)$ is both
small and completely determined by local conditions according to a
wonderfully convergent expansion (\ref{initial}). That is evident from
Figure~\ref{jump} even for an instantaneous jump in $\epsilon(t)$.

As long as $\partial_N h(t,k)$ remains small the full equation 
(\ref{finaleqn}) can be linearized to a form (\ref{lineqn}) for which we 
were able to derive a Green's function solution (\ref{GFsol2}). It cannot 
be overstressed that this solution pertains for an {\it arbitrary} 
inflationary expansion history. The assumption of linearity on which it 
is based should be valid long before first horizon crossing. It can break 
down long after first horizon crossing but in a way which is very simple 
to repair.

Our formalism differs from the generalized slow roll approximation
\cite{Stewart:2001cd,Dvorkin:2009ne} in three ways:
\begin{enumerate}
\item{Instead of correcting the mode function $u(t,k)$ and then inferring
how this affects $M(t,k) \equiv \vert u(t,k)\vert^2$, we correct $M(t,k)$
directly;}
\item{Our 0th order term is exact for arbitrary constant $\epsilon(t)$;
and}
\item{Our corrections are multiplicative rather than additive.}
\end{enumerate}
As the early time expansions (\ref{badcomp}) and (\ref{Mexp}) show, our 
formalism captures effects at first order which require going to second 
order in the generalized slow roll expansion. Given a specific model, the 
additional accuracy of our formalism is not required for the analysis of 
current data. Its advantage derives rather from the more explicit connection 
it makes between data and a general, initially unknown model. This has 
potential applications for the power spectra on three time scales:
\begin{enumerate}
\item{For current data it facilitates the process of inferring the sorts 
of models which might explain anomalies;}
\item{For next generation data, which might begin resolving the tensor 
power spectrum, it permits exploitation of the general relation (\ref{MtoN}) 
between the tensor and scalar power spectra to develop a version of the 
single scalar consistency relation \cite{Polarski:1995zn,GarciaBellido:1995fz,
Sasaki:1995aw} that could be employed before the tensor spectral index has 
been well measured; and}
\item{For far future data, when the full development of 21 cm cosmology 
might permit loop corrections to be resolved, it elucidates both when the 
loop counting parameter of $G H^2(t)$ should be evaluated, and whether or 
not there can be enhancements of the form 
$\epsilon_{\rm late}/\epsilon_{\rm early}$.}
\end{enumerate}

Our work also has three more general applications. First, there is a 
close relation between $M(t,k)$ and the vacuum expectation value of a 
massless, minimally coupled (MMC) scalar,
\begin{equation}
\Bigl\langle \Omega \Bigl\vert \varphi^2(t,\vec{x}) \Bigr\vert 
\Omega \Bigr\rangle = \int \!\! \frac{dk k^2}{2 \pi^2} \, M(t,k) \; .
\end{equation}
This relation should allow us to estimate the secular growth for an
arbitrary inflationary expansion history, which is an important step
in building nonlocal models to represent the quantum gravitational
back-reaction on inflation \cite{Tsamis:2009ja,Tsamis:2010ph,
Romania:2012av,Tsamis:2014hra}. Second, note that our transformation 
(\ref{MtoN}) could be used to convert the propagator of a MMC scalar 
into the propagator for the scalar perturbation field $\zeta(t,\vec{x})$ 
for an arbitrary inflationary expansion history. Of course we do not 
have MMC scalar propagator for arbitrary $a(t)$ but perhaps the 
transformation could be used to derive relations between loops involving 
gravitons and loops involving $\zeta$. Finally, our technique for 
passing from the oscillatory mode functions to their norm-squared
\cite{Romania:2011ez,Romania:2012tb} can be applied for any perturbations 
whose mode functions obey second order equations. It would be interesting 
to see what it gives for Higgs inflation and for $f(R)$ models of 
inflation.

\vskip 1cm

\centerline{\bf Acknowledgements}

We are grateful for conversations and correspondence on this subject 
with P. K. S. Dunsby, S. Odintsov, L. Patino, M. Romania, S. Shandera 
and M. Sloth. This work was partially supported by the 
European Union (European Social Fund, ESF) and Hellenic national funds 
through the Operational Program ``Education and Lifelong Learning" 
of the National Strategic Reference Framework (NSRF) under the 
``$\Theta\alpha\lambda\acute{\eta}\varsigma$'' action MIS-375734, 
under the ``$A\rho\iota\sigma\tau\epsilon\acute{\iota}\alpha$'' 
action, under the ``Funding of proposals that have received a 
positive evaluation in the 3rd and 4th Call of ERC Grant Schemes''; 
by NSF grants PHY-1205591 and PHY-1506513, and by the Institute for 
Fundamental Theory at the University of Florida.

\section{Appendix A: Simplifying Equation (\ref{DMeqn1})}

The first time derivative of $M_0(t,k)$ is,
\begin{equation}
\dot{M}_0 = -2 H M_0 + \dot{z} M_0' + \dot{\nu} \partial_{\nu} M_0 \; .
\label{M0dot}
\end{equation}
where a prime stands for the derivative with respect to $z$ and 
$\partial_{\nu}$ denotes differentiation with respect to $\nu$. 
It is best to postpone using the explicit expressions for $\dot{z}$ 
and $\dot{\nu}$,
\begin{equation}
\dot{z} = -\frac{k}{a} + \frac{\dot{\epsilon}}{1 \!-\! \epsilon}
\times z \qquad , \qquad \dot{\nu} = \frac{\dot{\epsilon}}{(1 \!-\! 
\epsilon)^2} \; . \label{edot}
\end{equation}
The time second derivative of $M_0(t,k)$ is,
\begin{eqnarray}
\lefteqn{\ddot{M}_0 = \Bigl(4 \!+\! 2 \epsilon\Bigr) H^2 M_0 + 
\Bigl(-4 H \dot{z} \!+\! \ddot{z}\Bigr) M_0' + \Bigl(-4 H \dot{\nu} 
\!+\! \ddot{\nu}\Bigr) \partial_{\nu} M_0 } \nonumber \\
& & \hspace{6cm} + \dot{z}^2 M_0'' + 2 \dot{z} \dot{\nu} 
\partial_{\nu} M_0' + \dot{\nu}^2 \partial_{\nu}^2 M_0 \; . \qquad
\label{M0ddot}
\end{eqnarray}
Bessel's equation implies that $M_0''$ can be eliminated using,
\begin{equation}
M_0'' + 2 M_0 - 2 (2 \!-\! \epsilon) \frac{H^2 a^2}{k^2} M_0 =
\frac1{2 M_0} \Bigl[ {M_0'}^2 + \frac1{k^2 a^4} \Bigr] \; . 
\label{M0eqn}
\end{equation}
The other derivatives we require are,
\begin{eqnarray}
3 H \dot{M}_0 & = & -6 H^2 M_0 + 3 H \dot{z} M_0' + 3 H \dot{\nu}
\partial_{\nu} M_0 \; , \qquad \\
-\frac{\dot{M}_0^2}{2 M_0} & = & -2 H^2 M_0 + 2 H \dot{z} M_0' +
2 H \dot{\nu} \partial_{\nu} M_0 - \frac{( \dot{z} M_0' \!+\! 
\dot{\nu} \partial_{\nu} M_0)^2}{2 M_0} \; . \qquad
\end{eqnarray}
Substituting everything into the definition (\ref{DMeqn1}) of 
$S(t,k)$ gives,
\begin{eqnarray}
\lefteqn{S(t,k) = -(H \dot{z} \!+\! \ddot{z}) \frac{M_0'}{M_0}
- (H \dot{\nu} \!+\! \ddot{\nu}) \frac{\partial_{\nu} M_0}{M_0}
- 2 \dot{z} \dot{\nu} \Bigl[ \frac{\partial_{\nu} M_0'}{M_0} \!-\! 
\frac{M_0'}{2 M_0} \frac{\partial_{\nu} M_0}{M_0} \Bigr] } 
\nonumber \\
& & \hspace{0cm} - \dot{\nu}^2 \Bigl[ \frac{\partial_{\nu}^2 M_0}{M_0}
\!-\! \frac12 \Bigl(\frac{\partial_{\nu} M_0}{M_0}\Bigr)^2 \Bigr] -
\Bigl( \dot{z}^2 \!-\! \frac{k^2}{a^2}\Bigr) \Biggl\{ \frac{(4 
\!-\! 2 \epsilon)}{(1 \!-\! \epsilon)^2 z^2} - 2 + \frac1{2 k^2 
a^4 M_0^2} \Biggr\} . \qquad \label{M0eqn2}
\end{eqnarray}

Each of the five terms on the right hand side of (\ref{M0eqn2}) is
proportional to at least one derivative of $\epsilon(t)$. Before
exhibiting this it is desirable to isolate the $\epsilon$-dependent
part of the index $\nu$, and to change from co-moving time $t$ to
the number of e-foldings since the beginning of inflation $N \equiv
\ln[a(t)/a_i]$,
\begin{equation}
\Delta \nu \equiv \frac1{1 \!-\! \epsilon} \qquad , \qquad
\frac{d}{d t} = H \frac{d}{d N} \quad , \quad \frac{d^2}{d t^2}
= H^2 \Bigl[ \frac{d^2}{d N^2} \!-\! \epsilon \frac{d}{d N} \Bigr]
\; .
\end{equation}
With this notation the five prefactors from the right hand side of 
(\ref{M0eqn2}) are, 
\begin{eqnarray}
H \dot{z} \!+\! \ddot{z} & \!\!\!\!\!=\!\!\!\!\! & 
\Bigl[ \frac{H \epsilon \dot{\epsilon}}{1 \!-\! \epsilon} \!+\! 
\frac{\ddot{\epsilon}}{1 \!-\! \epsilon}  \!+\! 2 \Bigl( 
\frac{ \dot{\epsilon}}{1 \!-\! \epsilon}\Bigr)^2 \Bigr] z = 
H^2 \Bigl[ \frac{\partial_N^2 \epsilon}{1 \!-\! \epsilon} \!+\! 
2 \Bigl( \frac{\partial_N \epsilon}{1 \!-\! \epsilon}\Bigr)^2
\Bigr] z \; , \qquad \label{pre1} \\
H \dot{\nu} \!+\! \ddot{\nu} & \!\!\!\!\!=\!\!\!\!\! & 
\Bigl[ \frac{H \dot{\epsilon}}{1 \!-\! \epsilon} \!+\! 
\frac{\ddot{\epsilon}}{1 \!-\! \epsilon} \!+\! 2 \Bigl( 
\frac{\dot{\epsilon}}{1 \!-\! \epsilon} \Bigr)^2 \Bigr] \Delta \nu 
\!=\! H^2 \Bigl[ \partial_N \epsilon \!+\! \frac{\partial_N^2 
\epsilon}{1 \!-\! \epsilon} \!+\! 2 \Bigl( \frac{\partial_N 
\epsilon}{1 \!-\! \epsilon}\Bigr)^2 \Bigr] \Delta \nu \; , \quad 
\label{pre2} \\
2 \dot{z} \dot{\nu} & \!\!\!\!\!=\!\!\!\!\! & 
\Bigl[-2 H \dot{\epsilon} + 2 \Bigl(\frac{\dot{\epsilon}}{1 
\!-\! \epsilon}\Bigr)^2 \Bigr] z \Delta \nu =
H^2 \Bigl[ -2\partial_N \epsilon \!+\! 2 \Bigl( \frac{\partial_N 
\epsilon}{1 \!-\! \epsilon}\Bigr)^2 \Bigr] z \Delta \nu \; , 
\label{pre3} \\
\dot{\nu}^2 & \!\!\!\!\!=\!\!\!\!\! & 
\Bigl( \frac{\dot{\epsilon}}{1 \!-\! \epsilon} \Bigr)^2 \Delta \nu^2 
= H^2 \Bigl( \frac{\partial_N \epsilon}{1 \!-\! \epsilon}\Bigr)^2
\Delta \nu^2 \; , \label{pre4} \\
\dot{z}^2 \!-\! \frac{k^2}{a^2} & \!\!\!\!\!=\!\!\!\!\! & 
\Bigl[-2 H \dot{\epsilon} \!+\! \Bigl( \frac{ \dot{\epsilon}}{1 \!-\! 
\epsilon} \Bigr)^2 \Bigr] z^2 = 
H^2 \Bigl[ -2 \partial_N \epsilon  \!+\! \Bigl( \frac{\partial_N 
\epsilon}{1 \!-\! \epsilon}\Bigr)^2 \Bigr] z^2 \; . \label{pre5}
\end{eqnarray}

In comparing expressions (\ref{pre1}-\ref{pre5}) with (\ref{M0eqn2})
it will be seen that every derivative with respect to $z$ is paired 
with one factor of $z$, and every derivative with respect to $\nu$ 
is paired with one factor of $\Delta \nu$. This suggests differentiating
with respect to the logarithms,
\begin{equation}
\zeta \equiv \ln(z) \quad \Longrightarrow \quad 
z \frac{\partial}{\partial z} = \frac{\partial}{\partial \zeta} \quad 
, \quad \xi \equiv \ln(\Delta \nu) \quad \Longrightarrow \quad
\Delta \nu \frac{\partial}{\partial \nu} = \frac{\partial}{\partial \xi}
\; . \label{zetaxi}
\end{equation}
From (\ref{M0eqn2}) it is also apparent that the factor of $2 k a^2$
in the denominator of $M_0(t,k)$ is always cancelled, either by ratios
or explicit factors. It is best to define a new variable for the
logarithm of the factor in the numerator (\ref{M0def}),
\begin{equation}
\sigma(z,\nu) \equiv \ln\Biggl[ \frac{\pi z}{2} \Bigl\vert 
H^{(1)}_{\nu}(z) \Bigr\vert^2\Biggr] = \ln\Biggl[ \frac{\pi z}{2}
\Bigl[ J_{\nu}^2(z) \!+\! N_{\nu}^2(z)\Bigr] \Biggr] \; . \label{sigma}
\end{equation}
Our final form for the source is therefore,
\begin{eqnarray}
\lefteqn{\frac{S(t,k)}{H^2} = -\Bigl[ \frac{\partial_N^2 
\epsilon}{1 \!-\! \epsilon} \!+\! 2 \Bigl( \frac{\partial_N
\epsilon}{1 \!-\! \epsilon}\Bigr)^2\Bigr] \frac{\partial \sigma}{
\partial \zeta} - \Bigl[ \partial_N \epsilon \!+\! 
\frac{\partial_N^2 \epsilon}{1 \!-\! \epsilon} \!+\! 2 \Bigl( 
\frac{\partial_N \epsilon}{1 \!-\! \epsilon}\Bigr)^2\Bigr] 
\frac{\partial \sigma}{\partial \xi} } \nonumber \\
& & \hspace{-.3cm} - 2 \Bigl[ -\partial_{N} \epsilon \!+\! 
\Bigl( \frac{\partial_N \epsilon}{1 \!-\! \epsilon} \Bigr)^2 \Bigr] 
\Bigl[ \frac{\partial^2 \sigma}{\partial \zeta \partial \xi} \!+\! 
\frac12 \frac{\partial \sigma}{\partial \zeta} \frac{\partial 
\sigma}{\partial \xi} \Bigr] - \Bigl( \frac{\partial_N \epsilon}{
1 \!-\! \epsilon}\Bigr)^2 \Bigl[\frac{\partial^2 \sigma}{\partial \xi^2} 
\!-\! \frac{\partial \sigma}{\partial \xi} \!+\! \frac12 \Bigl( 
\frac{\partial \sigma}{\partial \xi}\Bigr)^2 \Bigr] \nonumber \\
& & \hspace{3.5cm} - 2 \Bigl[ -2\partial_N \epsilon \!+\! \Bigl( 
\frac{\partial_N \epsilon}{1 \!-\! \epsilon} \Bigr)^2 \Bigr] 
\Bigl[ \frac{(2 \!-\! \epsilon)}{(1 \!-\! \epsilon)^2} + e^{2 \zeta} 
(e^{-2\sigma} \!-\! 1)\Bigr] \; . \label{M0eqn3}
\end{eqnarray}

We obviously want to make the same changes on the left hand side of
(\ref{M0eqn2}). It is also desirable to change the dependent variable
from $\Delta M$ to $h(t,k) \equiv -2\ln[\Delta M(t,k)]$, all of which 
implies,
\begin{eqnarray}
\lefteqn{\frac{\Delta \ddot{M}}{\Delta M} + \Bigl[ 3 H + 
\frac{\dot{M}_0}{M_0} \Bigr] \frac{\Delta \dot{M}}{\Delta M} - \frac12 
\Bigl(\frac{ \Delta \dot{M}}{\Delta M}\Bigr)^2 + \frac1{2 a^6 M_0^2} 
\Bigl[1 \!-\! \frac1{\Delta M^2} \Bigr] \Biggr] } \nonumber \\
& & \hspace{-.5cm} = -\frac{H^2}{2} \Biggl\{ \partial^2_N h \!-\! 
\Bigl[\frac12 \partial_N h\Bigr]^2 \!+\! \Bigl[1 \!-\! \epsilon \!+\! 
\partial_N \sigma\Bigr] \partial_N h \!+\! \Bigl[2 (1 \!-\! \epsilon) 
e^{\zeta - \sigma}\Bigr]^2 \Bigl[e^{h} \!-\! 1\Bigr] \Biggr\} . 
\qquad \label{DMeqn4}
\end{eqnarray}
Equating (\ref{M0eqn3}) to (\ref{DMeqn4}) and dividing out the common
factor of $-\frac12 H^2$ gives the final form (\ref{finaleqn}) of our
evolution equation.

\section{Appendix B: Equation (\ref{finaleqn}) at Early Times}

In the early time regime the parameter $z(t,k) \equiv \frac{k}{
(1-\epsilon)Ha}$ is large which implies,
\begin{equation}
\frac{\pi z}{2} \Bigl\vert H^{(1)}_{\nu}(z)\Bigr\vert^2 = 1 + 
\frac{(\nu^2 \!-\! \frac14)}{2 z^2} + \frac{3(\nu^2 \!-\! \frac14) 
(\nu^2 \!-\! \frac94)}{8 z^4} + O\Bigl(\frac1{z^6}\Bigr) \; .
\end{equation}
Hence the early time expansion for $\sigma(z,\nu)$ is,
\begin{equation}
\sigma(z,\nu) = \frac{(\nu^2 \!-\! \frac14)}{2 z^2} +
\frac{(\nu^2 \!-\! \frac14) (\nu^2 \!-\! \frac{13}4)}{4 z^4}
+ \dots \label{sigbigz}
\end{equation} 
Expression (\ref{sigbigz}) implies the following expansions for the 
various $\sigma$-dependent factors in (\ref{finaleqn}),
\begin{eqnarray}
\Bigl[1 \!-\! \epsilon \!+\! \partial_N \sigma\Bigr] & = &
1 \!-\! \epsilon + O\Bigl(\frac1{z^2}\Bigr) \; , \\
\Bigl[2 (1 \!-\! \epsilon) e^{\zeta - \sigma}\Bigr]^2 & = &
4 (1 \!-\! \epsilon)^2 z^2 + O(1) \; , \\
\frac{\partial \sigma}{\partial \zeta} = -\frac{(\nu^2 
\!-\! \frac14)}{z^2} + O\Bigl(\frac1{z^4}\Bigr)  & , &  
\frac{\partial \sigma}{\partial \xi} = \frac{(\nu^2 
\!-\! \frac12 \nu)}{z^2} + O\Bigl(\frac1{z^4}\Bigr) \; , \\
\Bigl[ \frac{\partial^2 \sigma}{\partial \zeta \partial \xi} \!+\! 
\frac12 \frac{\partial \sigma}{\partial \zeta} \frac{\partial 
\sigma}{\partial \xi} \Bigr] & = & -\frac{(2\nu^2 \!-\! \nu)}{z^2} 
+ O\Bigl(\frac1{z^4}\Bigr) \; , \\
\Bigl[\frac{\partial^2 \sigma}{\partial \xi^2} \!-\! \frac{\partial 
\sigma}{\partial \xi} \!+\! \frac12 \Bigl( \frac{\partial \sigma}{
\partial \xi}\Bigr)^2 \Bigr] & = & \frac{(\nu \!-\! \frac12)^2}{z^2} 
+ O\Bigl(\frac1{z^4}\Bigr) \; , \\
\Bigl[ \frac{(2 \!-\! \epsilon)}{(1 \!-\! \epsilon)^2} + 
e^{2 \zeta} (e^{-2\sigma} \!-\! 1)\Bigr] & = & \frac{3 (\nu^2 
\!-\! \frac14)}{2 z^2} + O\Bigl( \frac1{z^4}\Bigr) \; .
\end{eqnarray}
Substituting these expansions in (\ref{finaleqn}) and additionally
neglecting nonlinear terms in $h(t,k)$ gives equation (\ref{early}).

\section{Appendix C: Equation (\ref{finaleqn}) at Late Times}

In the late time regime of $z(t,k) \ll 1$, but still with 
$0 \leq \epsilon(t) < 1$, it is the small argument expansion of the
Neumann function in (\ref{sigma}) which controls the behavior of 
$\sigma(z,\nu)$,
\begin{equation}
\sigma = \ln\Biggl[ \frac{\Gamma^2(\nu)}{\pi}
\Bigl( \frac{2}{z}\Bigr)^{2\nu - 1} \Biggr] + O(z^2) =
2 \Delta \nu \Biggl[ \Delta N \!+\! \ln\Bigl[ \frac{H}{H(t_k)}\Bigr]
\Biggr] + \ln\Bigl[ C(\epsilon)\Bigr] + O(z^2) \; . \label{smallsig} 
\end{equation}
Its derivative involves the digamma function $\psi(z) \equiv 
\frac{d}{dz} \ln[\Gamma(z)]$,
\begin{equation}
\partial_N \sigma = 2 + 2 \Delta \nu \frac{\partial_N \epsilon}{1
\!-\! \epsilon} \Bigl[ -1 \!-\! \ln\Bigl(\frac{z}{2}\Bigr) \!+\! 
\psi\Bigl( \frac12 \!+\! \Delta \nu\Bigr) \Bigr] + O(z^2) \; .
\end{equation}
The term in square brackets is defined in expression (\ref{Fdef}) 
and grows roughly linearly in $N$. The seven $\sigma$-dependent 
factors of expression (\ref{finaleqn}) have the expansions,
\begin{eqnarray}
\Bigl[1 \!-\! \epsilon \!+\! \partial_N \sigma\Bigr] & = &
3 \!-\! \epsilon + \frac{\partial_N \epsilon}{1 \!-\! \epsilon}
\times 2 \Delta \nu F + O(z^2) \; , \label{late1} \\
\Bigl[2 (1 \!-\! \epsilon) e^{\zeta - \sigma}\Bigr]^2 & = &
4 (1\!-\!\epsilon)^2 z^2 \times \Bigl( \frac{z}{2}\Bigr)^{4
\Delta \nu} \frac{\pi^2}{\Gamma^4(\nu)} + O(z^{4 + 4 \Delta \nu}) 
\; , \qquad \label{late2} \\
\frac{\partial \sigma}{\partial \zeta} = - 2\Delta \nu +
O(z^2) & , &
\frac{\partial \sigma}{\partial \xi} = 2 \Delta \nu (F \!+\!1)
+ O(z^2) \; , \label{late34} \\
\Bigl[ \frac{\partial^2 \sigma}{\partial \zeta \partial \xi} \!+\! 
\frac12 \frac{\partial \sigma}{\partial \zeta} \frac{\partial 
\sigma}{\partial \xi} \Bigr] & = & - 2\Delta \nu^2 (F \!+\! 1) -
2 \Delta \nu + O(z^2) \; , \label{late5} \\
\Bigl[\frac{\partial^2 \sigma}{\partial \xi^2} \!-\! \frac{\partial 
\sigma}{\partial \xi} \!+\! \frac12 \Bigl( \frac{\partial \sigma}{
\partial \xi}\Bigr)^2 \Bigr] & = & 2 \Delta \nu^2 \Bigl[ (F \!+\! 
1)^2 + \psi'\Bigl( \frac12 \!+\! \Delta \nu\Bigr)\Bigr] + O(z^2)
\; , \label{late6} \\
\Bigl[ \frac{(2 \!-\! \epsilon)}{(1 \!-\! \epsilon)^2} + 
e^{2 \zeta} (e^{-2\sigma} \!-\! 1)\Bigr] & = & \Delta \nu + 
\Delta \nu^2 + O(z^2) \label{late7} \; .
\end{eqnarray}
From (\ref{late2}) the restoring force drops out of (\ref{finaleqn}) 
but all the other terms contribute to give the late time limiting
form (\ref{lateqn}).

\end{document}